\documentclass[amsmath,amssymb,aps, prb,twocolumn,groupedaddress,notitlepage]{revtex4-2}

\usepackage{graphicx}
\usepackage[normalem]{ulem}
\usepackage{xcolor}
\begin{document}
\preprint{APS/123-QED}
\newcommand{\bto}{BaTi$\text{O}_3$ }
\newcommand{\red}[1]{\textcolor{red}{#1}}
\newcommand{\blue}[1]{\textcolor{blue}{#1}}

% \title{Ab initio modeling of ferroelectrics phase transition: The case of PbTi$\text{O}_3$}
% \title{Reconcile displacive softening and strong disordering for ferroelectric phase transition: the case of lead titanate}

\title{Thermal disorder and phonon softening in the ferroelectric phase transition of lead titanate}
 
\author{Pinchen Xie}
\thanks{Current address: Applied Mathematics and Computational Research Division, Lawrence Berkeley National Laboratory, Berkeley, CA 94720, USA}
\author{Yixiao Chen}
\affiliation {Program in Applied and Computational Mathematics, Princeton University, Princeton, NJ 08544, USA}
\author{Weinan E}
\affiliation{AI for Science Institute, Beijing, China,\\
Center for Machine Learning Research and School of Mathematical Sciences, Peking University, Beijing, China}
\author{Roberto Car}
\thanks{rcar@princeton.edu}
\affiliation {Department of Chemistry, Department of Physics, Program in Applied and Computational Mathematics, Princeton Materials Institute, Princeton University, Princeton, NJ 08544, USA}

\date{\today}
\begin{abstract}
    We report a molecular dynamics study of ab initio quality of the ferroelectric phase transition in crystalline PbTi$\text{O}_3$. We model anharmonicity accurately in terms of potential energy and polarization surfaces trained on density functional theory data with modern machine learning techniques. Our simulations demonstrate that the transition has a strong order-disorder character, in agreement with diffraction experiments, and provide fresh insight into the approach to equilibrium across the phase transition. We find that the emergence and disappearance of the macroscopic polarization is driven by dipolar switching at the nanometer scale. We also computed the infrared optical absorption spectra in both the ferroelectric and the paraelectric phases, finding good agreement with the experimental Raman frequencies. Often, the almost ideal displacive character of the soft mode detected by Raman scattering in the paraelectric phase has been contrasted with the order-disorder character of the transition suggested by diffraction experiments. We settle this issue by showing that the soft mode coexists with a strong Debye relaxation associated with thermal disordering of the dipoles. The Debye relaxation feature is centered at zero frequency and appears near the transition temperature in both the ferroelectric and the paraelectric phases.  
     
\end{abstract}

\maketitle

\section{Introduction}
Over the past decades, the focus of the community interested in ferroelectricity has evolved from 
conventional bulk ferroelectrics 
to relaxors~\cite{bokov2020}, multiferroics~\cite{spaldin2010multiferroics}, two-dimensional materials~\cite{guan2020recent}, and ferroelectric polymers~\cite{lovinger1983ferroelectric}. 
However, some issues pertaining to the dynamics of the ferroelectric phase transition, notably how equilibrium is approached after an abrupt change in thermodynamic conditions or how spectral changes across the transition relate to the atomic motion, have not been fully elucidated even in the context of simple bulk materials. Here, we address these issues in the case of the prototypical ferroelectric crystal PbTi$\text{O}_3$.  

In this system, the change of static equilibrium properties across the ferroelectric to paraelectric (FE-PE) phase transition is well understood. The spontaneous polarization of the FE phase vanishes in the PE phase because of the loss of long-range order among local polar distortions that persist and are not significantly weaker in the PE phase. This mechanism, called order-disorder, is supported by experiments~\cite{nelmes1990order, ravel1993lead,sicron1994nature,ravel1995order, chapman2005diffuse, datta2018stochastic} and model calculations based on physically motivated force fields~\cite{Shin_2008}. A different mechanism, called displacive, was often invoked in the early studies of ferroelectricity. It stipulates that the transition is caused by a structural instability of the lattice, associated with the softening of the slowest optical phonon of frequency $\omega_s$. In this scheme, when the temperature $T$ crosses the transition temperature $T_c$, the lattice deforms uniformly and irreversibly, while $\omega^2_s(T)$ decays linearly with the temperature as $T$ approaches $T_c$ on either side~\cite{venkataraman1979soft}. In current understanding, order-disorder and displacive mechanisms coexist in most ferroelectric materials as a consequence of anharmonicity. 

The interpretation of the dynamics across the phase transition in PbTi$\text{O}_3$ is less straightforward. On the theoretical side, calculations have been limited to approximate treatments of anharmonicity and its temperature dependence. On the experimental side, real-time measurements of the atomic motions have not been accessible on the picosecond time scale. Thus, we do not know exactly how a sample subjected to a rapid temperature change across $T_c$ would evolve towards a new equilibrium state. 
Most available experimental information is spectroscopic, i.e., time-integrated.    
In PbTi$\text{O}_3$, Raman spectra show that,
when $T_c$ is approached from below, $\omega^2_s(T)$ decays to zero faster than linearly~\cite{fontana1990, fontana1991}, 
a behavior attributed to a crossover of the dominant microscopic mechanism, from displacive to order-disorder, at a temperature of more than one hundred degrees below $T_c$~\cite{fontana1990}. The importance of disordering is further supported by the finding of a central quasi-elastic peak in Raman scattering experiments, distinct from the sharper elastic peak, both below and above $T_c$~\cite{fontana1990}. The quasi-elastic peak originates from the Debye relaxation of the local dipoles as a result of disordering fluctuations.
Surprisingly, multiple light and neutron scattering experiments indicate that, when $T_c$ is approached from above, the slowest optical mode behaves as expected for a displacive transition, i.e., it is not significantly overdamped and decays to zero linearly with $T-T_c$ in a nearly ideal way~\cite{Shirane1970, Tomeno2012, Hlinka2013}. 
This appears counterintuitive, since disordering should unravel the spatial and temporal coherence of the slowest zone-center optical mode, significantly weakening its displacive features. An extreme case of this occurs in relaxor ferroelectrics like $\rm{Pb\left(Mg_{1/3}Nb_{2/3}\right)O_3}$, where disordered polar nanodomains overdamp the soft mode so much that detecting a linear temperature dependence of $\omega^2_s(T)$ becomes meaningless~\cite{wakimoto2002ferroelectric}.

In this paper, we address the above challenges by modeling PbTi$\text{O}_3$ with DFT precision all-atom molecular dynamics simulations, made possible by machine learning. We use two atomistic neural-network models, a deep potential (DP) model~\cite{PhysRevLett.120.143001, zhang2018end}  representing the Born-Oppenheimer potential energy surface and a Deep Dipole (DD) model~\cite{PhysRevB.102.041121} representing the polarization surface and its decomposition into local dipoles ~\cite{vanderbilt2018berry, resta2007theory}. 
Our approach overcomes the limitations of effective Hamiltonian theories~\cite{Zhong1994,zhong1995,PhysRevB.55.6161, PhysRevB.78.104104} in capturing strong anharmonicity~\cite{Tinte2003}, as well as the limitations of hand-crafted force fields ~\cite{Tinte1999, Goddard2002, grinberg2002, Brown2009, PhysRevB.88.104102} in modeling the potential energy surface with uniform accuracy in the thermodynamic range of interest.
    
After an artificial hydrostatic pressure was imposed on the system to correct for the supertetragonality of the adopted DFT approximation, DP and DD models predict equilibrium structural and dielectric properties in good agreement with experiments across the FE-PE phase transition. To gain insight into the dynamics of disorder in real time, we study how long-range order among the local dipoles is lost or acquired
when the system is brought out of equilibrium by an instantaneous change of temperature across $T_c$.  We find that equilibrium is restored by nanoscale fluctuations that are uniformly distributed in space rather than by the formation of a growing droplet of the new equilibrium phase embedded in the old phase, as in the case of a typical first-order transition like melting. This likely occurs because in the present case, the macroscopic coexistence between a FE and a PE crystal is hampered by their different lattice constants. However, we cannot exclude that macroscopic coexistence could occur in the thermodynamic limit. Next, we study the change with temperature of the infrared (IR) absorption coefficient within the linear response theory. The simulated IR features have frequencies that closely agree with the experimental Raman spectra, as expected because
the IR active modes in PbTi$\text{O}_3$ are also Raman active. In agreement with experiments, we also find that, when $T_c$ is approached from below, $\omega_s^2(T)$ decays to zero faster than predicted by the ``soft mode'' theory, but it decays linearly with $T$, when $T_c$ is approached from above. To reconcile the last finding with the dominant order-disorder mechanism of the transition, we observe that the IR spectral function contains an $\omega^2$ factor that suppresses the IR response in the vicinity of $\omega=0$. 
To eliminate this effect, we consider a spectral response function without the $\omega^2$ factor, finding, below and above $T_c$, a Debye relaxation mode centered at $\omega=0$. %, in good agreement with Ref.~\cite{fontana1990}. 
 The relaxation mode becomes broad in the paraelectric phase and contributes to the formation of the ``soft mode'' found above $T_c$.  Thus, the picture of a transition dominated by order disorder is corroborated not only by static but also by dynamic equilibrium properties.   

The paper is organized as follows. 

In Sec.~\ref{sec-md} we summarize the adopted DP and DD theoretical models and report the results of MD simulations for the lattice parameters, the enthalpy, the spontaneous polarization, the specific heat, and the dielectric susceptibility. Our results agree well with experiments across the FE-PE transition. 

In Sect.~\ref{sec-disorder}, we provide direct evidence for the strong order-disorder character of the FE-PE transition in terms of the one- and two-body distributions of the local dipole moments associated with the elementary cells of the crystal. In addition, we used nonequilibrium MD to compute FE-PE phase-transition trajectories, finding that the disorder develops by stochastic excitation of polar and nonpolar nanoregions without a nucleation stage.  

In Sect.~\ref{sec-ir} we contrast the strong disorder of the paraelectric phase with the presence, for $T>T_c$, of a non-overdamped ``soft mode'' in the far-infrared absorption spectra. To understand the origin of this mode, we examine the power spectrum of the total dipole moment, since the corresponding spectral function does not include the $\omega^2$ factor present in the IR spectra. We find that a zero frequency Debye relaxation mode due to disorder is present near $T_c$, both below and above the transition, in good agreement with the experiments reported in Ref.~\cite{fontana1990}. 
% Furthermore, a zero-frequency mode is also found to coexist with the optical mode in the power spectrum of polarization for $T<T_c$. Such zero-frequency mode is consistent with the central peak found in

Finally, Sect.~\ref{sec:conclusion} is devoted to our conclusions.  
%, where we make further connections of our results to the displacive to order-disorder crossover suggested by Ref.~\cite{fontana1990} for the phase transition of PbTi$\text{O}_3$. }

% The spatial and time coherence of the latter should be largely suppressed in the paraelectric phase.  
% The presumed competition between strong disordering and ideal displacive softening is hence reconciled. 
 
\section{DFT-based atomistic models and equilibrium properties}\label{sec-md}
 
We adopt the strongly constrained and appropriately normed meta-GGA functional approximation (SCAN)~\cite{sun2015strongly} of DFT. 
%It is shown in Ref.~\cite{scan2017} that SCAN-DFT systematically improves the prediction of structural and electric properties of a wide class of ferroelectric materials compared to other general-purpose functional approximations. 
The details of our SCAN-DFT calculations and the corresponding properties of the classical ground-state structure of PbTi$\text{O}_3$ are reported in the Supplemental Material~\cite{SupplementalMaterial}. 
%%%%%%%%%%%%%%%%%
\begin{figure}[th]
    \centering
    \includegraphics[width=0.9\linewidth]{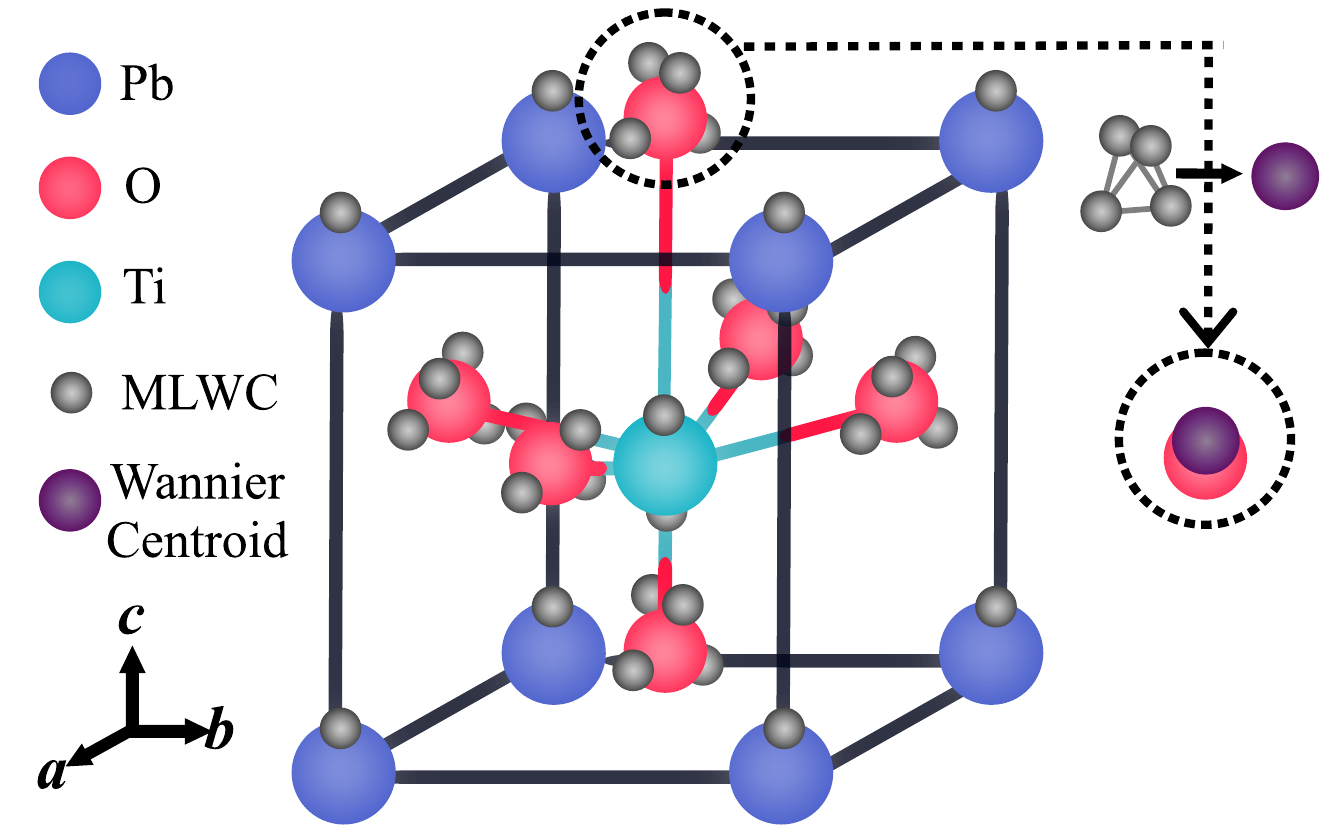}
    \caption{Conventional tetragonal cell for the ground state P4mm structure of PbTi$\text{O}_3$. Valence Wannier centers are indicated by grey spheres. Semicore Wannier centers for Ti and Pb are not shown. The Wannier centroid (purple) is the spatial average of the Wannier centers associated with the same atom, as illustrated for the topmost oxygen atom.}
    \label{wcs}
\end{figure}
%%%%%%%%%%%%%%%%%
% Detailed studies of the ground-state properties of PbTi$\text{O}_3$ based on the SCAN functional have been reported in the literature ~\cite{Paul2017,  scan2017}, using electronic structure calculations with the projector augmented wave (PAW) method  ~\cite{blochl1994projector}. Here, we use plane waves and norm-conserving pseudopotentials (NCPP)~\cite{hamann1979norm} instead. Our results for the static properties at zero temperature are reported in Appendix~\ref{app-dft}: they agree well with prior work, but minor differences are to be expected from the distinct numerical methods and convergence criteria adopted.
Fig.~\ref{wcs} shows the P4mm ground state structure of a Ti-centered primitive cell of PbTi$\text{O}_3$, including the centers of the maximally localized Wannier distribution~\cite{marzari2012maximally}. Within DFT, the Wannier centers associated with any atomic configuration of a periodic supercell, a multiple of the primitive cell, 
are obtained with a unitary transformation of the occupied Kohn-Sham orbital subspace~\cite{marzari2012maximally}. In PbTi$\text{O}_3$ each Wannier center can be uniquely assigned to its nearest atom and, due to the absence of electron transfer processes, this correspondence is preserved along the molecular dynamics trajectories. Hereafter, the geometric center of the Wannier centers assigned to a given atom will be called the Wannier centroid (WC)~\cite{dplr}. The WCs provide an effective point-charge representation of the valence electrons. The WC associated with atom-$i$ has position $W_i$ and its charge $q_i$ is equal to the total charge of its parent Wannier centers. The ion core of the atom $i$ has position $R_i$ and charge $Q_i$. A local dipole moment~\cite{meyer2002ab}, $p_j$, can then be assigned to each Ti-centered primitive cell:
\begin{equation}
    p_j = \sum_{i\sim j} \alpha_iQ_i d(R_i,R_j) + \alpha_i q_id(W_i,R_j).
\end{equation}
Here, $j$ labels a central Ti atom and a primitive cell, the summation is over the eight Pb atoms associated to that cell with weight $\alpha_i=1/8$, the six O atoms associated to that cell with weight $\alpha_i=1/2$, and the central Ti atom with weight $\alpha_i=1$ (see Fig.~\ref{wcs}), $d(R_i,R_j)=R_i-R_j$ and, similarly, $d(W_i,R_j)=W_i-R_j$, under minimum image convention. Since $p_j$ vanishes for a centrosymmetric unit cell, the centrosymmetric structure is taken as the reference for the zero of the polarization, fixing in this way the gauge freedom. The total dipole moment of a supercell of volume $V$ is $p^G=\sum_j p_j$. The corresponding polarization~\cite{resta2007theory} is $\mathcal{P}=p^G/V$.  

\begin{figure*}
    \centering
    \includegraphics[width=\linewidth]{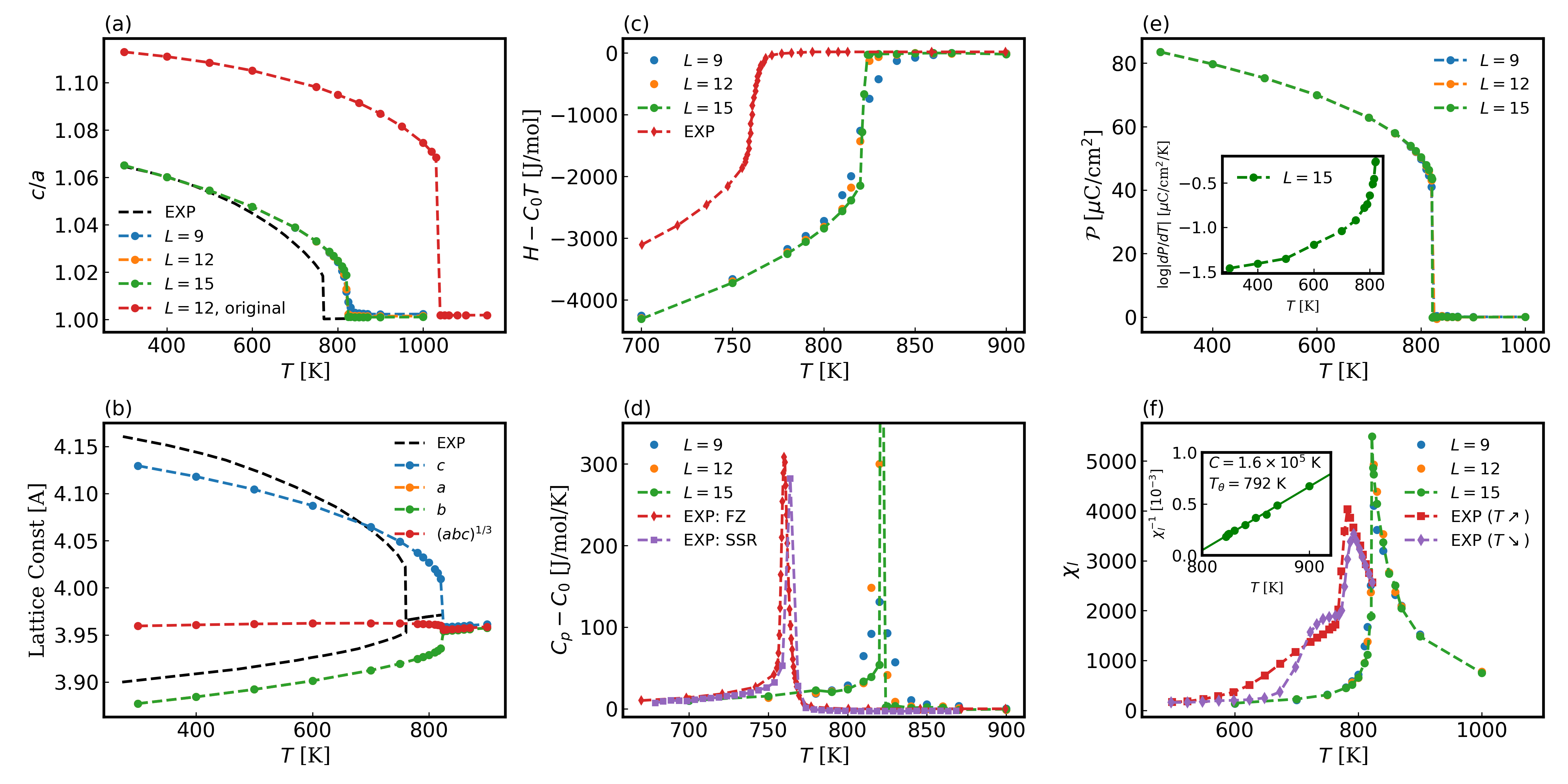}
    \caption{(a) The tetragonality of bulk PbTi$\text{O}_3$ for NPT-ensemble with $P=P_0$ (purple) and $P=P_a+P_0$ (others). The experimental data are excerpted from ~\cite{Mabud1979}. (b) The lattice constants of bulk PbTi$\text{O}_3$ for NPT-ensemble with $P=P_a+P_0$. The orange line is overlapping with the green line thus invisible. The red line represents $\Omega^{1/3}=(abc)^{1/3}$. (c) The difference per mole between the enthalpy $H$ and $C_0T=3nRT$. $n=5$ is the number of atoms in one unit of PbTi$\text{O}_3$. The curves are shifted along the vertical axis for easier comparison. (d) The difference between the computed specific heat $C_p$ and the Dulong-Petit specific heat $C_0=3nR$ for bulk PbTi$\text{O}_3$. 
    The experimental data are from ~\cite{Rossetti_2005} (red) and ~\cite{yoshida2009heat} (purple). (e) The temperature dependence of the spontaneous polarization for finite size systems $L=9,12,15$. The inset shows the computed pyroelectric coefficient for the ferroelectric region. (f) The longitudinal susceptibility $\chi_l$ near transition. The experimental data taken from the heating ($T \nearrow $) and cooling ($T \searrow $) cycles are excerpted from ~\cite{maffei_rossetti_2004}. The inset shows the Curie-Weiss behavior of $\chi_l^{-1}$ at the cubic phase. The guiding line is fitted to the data points (solid circle) for $L=15$.}
    \label{figmd}
\end{figure*}

% \twocolumngrid
%%%%%%%%%%%%%%%%%%%%%%%%%%%%%%%%%%%%%%%% 
As detailed in the Supplemental Material~\cite{SupplementalMaterial}, we train a DP and a DD model with DFT data for the potential energy surface, atomic forces, local dipole moments, and the polarization surface $\mathcal{P}$, using an active learning protocol~\cite{zhang2019active}, which was fine-tuned to generate uniformly accurate DP and DD models in the thermodynamic range of interest, that is, $T\in [300, 1200]K$ and pressure $P\in [0, 10^5]\text{Pa}$. In both the DP and the DD models the dependence on the atomic environment is limited by a spherical cutoff radius of $6$\r{A}. In the following, the DP and DD models will also be called the energy and dipole models, respectively. The root mean square error of the DP model in the training set was $0.7$meV per atom for energy prediction and $0.29\mathrm{eV/\AA}$ per atom for force prediction, while the validation errors in a representative set of configurations independent of the training set were $1.0$meV per atom for energy prediction and $0.35\mathrm{eV/\AA}$ per atom for force prediction. For the DD model, the root mean square error for polarization prediction amounts to $1.1\mathrm{\mu C/cm^2}$ on the training set and to $1.4\mathrm{\mu C/cm^2}$ on the validation set.
% Typical errors of 1 meV/atom and of $ 1 \mu \text{C/cm}^2$, are found for the energy and the polarization, respectively. 

The energy model is an accurate representation of SCAN-DFT. Like many other functional approximations at the GGA or metaGGA level, SCAN-DFT overestimates the tetragonality $c/a$ of PbTi$\text{O}_3$~\cite{scan2017}. 
The extent of this error is evident in panel (a) of Fig.~\ref{figmd}, in which the experimental ratio $c/a$ is compared with the results of NPT-MD simulations at ambient pressure ($P_0$). The large tetragonality of the theoretical model correlates with an overestimation of $T_c$ by almost $300$K.
As suggested in~\cite{Zhong1994}, the tetragonality error can be corrected to a large extent by adding an artificial hydrostatic pressure $P_a$ to the pressure $P_0$ acting on the theoretical sample. Here, we adopt $P_a=2.8\times 10^4$bar by matching theoretical and experimental tetragonalities at room temperature ($T=300K$). With this simple fix, the predicted ratio $c/a$ is much closer to the experiment in the temperature range of interest. In Fig.~\ref{figmd}(a), the effects of finite size on $c/a$ are examined for different supercell sizes, indicated by $L\times L \times L$ in units of the elementary cell. %Clearly, the additional pressure brings the tetragonality and also the transition temperature of the model in much closer agreement with the experiment.
% The plots for different cell sizes in Fig.~\ref{figmd} (a) illustrate the finite size dependence of the transition, which becomes sharper with increasing size, a behavior consistent with a first-order character of the transition. 
Under $P=P_a+P_0$, we find that the lattice constants shown in Fig.~\ref{figmd}(b) agree well with the experiment~\cite{Mabud1979} over the entire temperature range. 
% In addition, the plot shows that the MD simulation captures well the small thermal expansion of the unit cell volume $\Omega=V/L^3$ in the cubic phase.
%In our simulations, we found this artificial pressure makes nearly all observables closer to the experiments under atmospheric pressure.
In the following, unless otherwise specified, all reported MD simulations are performed in the NPT ensemble with $P=P_a+P_0$.

Next, we consider the thermodynamic properties of the bulk PbTi$\text{O}_3$. Fig.~\ref{figmd} (c) shows the temperature dependence of the enthalpy measured in experiments and in simulations on three different supercells, relative to the prediction by the law of Dulong and Petit. In the simulations, the enthalpy $H$ is computed from the NPT ensemble average of $E+P_0V$, where $E$ is the internal energy of the system. The experimental data were measured on single bulk crystals grown by the floating zone technique (FZ)~\cite{Rossetti_2005}. 
In simulations, finite-size effects are small when $L\geqslant 12$. The simulations yield a slightly overestimated latent heat of approximately $2000$ J / mol compared to the experiment of~\cite{Rossetti_2005}. %Both in the experiment and in simulations the cubic phase satisfies well the equipartition law. 
% The experimental latent heat estimated from Fig.~\ref{figmd} (c) is only slightly smaller.
In Fig.~\ref{figmd} (d), the excess specific heat $C_p-C_0$ obtained from the simulations is compared to the experimental results of Ref.~\cite{Rossetti_2005} and Ref.~\cite{yoshida2009heat}.   
In the simulations, $C_p$ is extracted from the fluctuation of the enthalpy $H$. $C_0=3nR$ is predicted by the Dulong-Petit law.  
% The bulk properties of float-zone samples should be closer to the properties of a pure single-domain crystal, than the properties of powder and flux-grown ceramic samples~\cite{bhide1968,REMEIKA197037}.
% Notably, the measured specific heat of flux-grown ceramic PbTi$\text{O}_3$ did not show Dulong-Petit-like behavior in the cubic phase, while our computation and the two experiments shown here obey closely the Dulong-Petit law in the cubic phase. 
% However, float-zone samples near the transition should still contain a considerable amount of defects.
In a small interval around the transition temperature, i.e., for $T=T_c\pm 5K$, the simulated peak of $C_p$ is narrower and sharper than in experiments, presumably due to defects present in the experimental samples. %The experimental observations may be affected by strain inhomogeneities \cite{Rossetti_2005} that changed the local transition temperature in the samples, resulting in an extended phase coexistence that smoothes out the singularity of the heat capacity.

Lastly, we consider dielectric properties. With our dipole model, these properties can be calculated with full inclusion of anharmonicity, in contrast to the dynamic Born charge approximation. 
Fig.~\ref{figmd} (e) shows the temperature dependence of spontaneous polarization $\mathcal{P}$. At $T=300$K,  $\mathcal{P}$ is equal to $84\mu \text{C}/\text{cm}^2$ and the pyroelectric coefficient ($d\mathcal{P}/{dT}$) is equal to $34\text{nC}\cdot\text{cm}^{-2}K^{-1}$.
In experiments, $\mathcal{P}$ varies from $70\mu \text{C}/\text{cm}^2$ to $100\mu \text{C}/\text{cm}^2$~\cite{nishino2020,Dahl2009,Morita2004}, and the 
pyroelectric coefficient varies from $24\text{nC}\cdot\text{cm}^{-2}K^{-1}$ to $27\text{nC}\cdot\text{cm}^{-2}K^{-1}$~\cite{Deb1995,Kenji1986}.
% So far, relatively accurate measurements of bulk $\mathcal{P}$ are obtained from extrapolation of thin films results. 
% The $\mathcal{P}$ values at room temperature, estimated in this way, vary from $70\mu \text{C}/\text{cm}^2$ to $100\mu \text{C}/\text{cm}^2$~\cite{nishino2020,Dahl2009,Morita2004}. The room temperature values for the pyroelectric coefficient estimated in experiments are in the range from $24\text{nC}\cdot\text{cm}^{-2}K^{-1}$ to $27\text{nC}\cdot\text{cm}^{-2}K^{-1}$~\cite{Deb1995,Kenji1986}. 
%Our simulation results are compatible with these experimental values.  
Fig.~\ref{figmd} (f) shows $\chi_l(T)$, the longitudinal zero-field static dielectric susceptibility of bulk PbTi$\text{O}_3$. In the simulations, well-converged $\chi_l(T)$ for different cell sizes are computed from the fluctuation of the polarization.  This calculation does not include the small contribution due to $\chi_\infty(T)$, the electronic contribution at clamped ions. The latter could be evaluated with a deep model for polarizability~\cite{sommers2020raman}.
$\chi_l(T)$ has a sharp peak near $T=821$ K, indicating a first-order ferroelectric transition. For comparison, the experimental $\chi_l(T)$ ~\cite{maffei_rossetti_2004} shows a shoulder at $T=763K$, the experimental phase transition temperature, and a broader peak shifted to temperatures closer to $T=793$K. This distortion has been attributed to domain pinning caused by defects and internal stresses in the sample~\cite{maffei_rossetti_2004}.
In contrast, all MD data points in the figure have been obtained from thermally equilibrated independent trajectories, as detailed in the Supplemental Material~\cite{SupplementalMaterial}.
The computed susceptibility allows the examination of the Curie-Weiss law. The inset of Fig.~\ref{figmd} (f) shows a very good linear temperature dependence of $\chi_l(T)^{-1}$ in the cubic phase. The optimized Curie constant and Curie temperature are $C=1.6\times 10^5$K and $T_\theta=792$K, respectively. 
The experimental values reported of $T_\theta$ are consistently close to the experimental phase transition temperature, but $C$ varies from $1.1\times 10^5$K to $4.1\times 10^5$K~\cite{bhide1968, Rossetti_2005,REMEIKA197037}, likely due to different concentrations of defects in the experimental samples.

In the effective Hamiltonian context, an accurate treatment of the dipole-dipole electrostatic interaction, including the power law decay at large separation distance, was deemed necessary to achieve agreement with experiment~\cite{zhong1995}. Interestingly, our energy model, which only includes finite range dipolar interactions, can capture well the singular behavior of $\chi_l$. Long-range electrostatics is responsible for the splitting of longitudinal (LO) and transverse (TO) optical phonons at long wavelengths in polar crystals and could be included in a DP model as suggested in Ref.~\cite{dplr}. However, this seems unnecessary because the LO-TO splitting gives only a minor contribution to $\chi_l$ near $T_c$ in PbTi$\text{O}_3$, as one can infer from the Lyddane-Sachs-Teller relation ${\epsilon_0}/{\epsilon_\infty}={\omega_{\mathrm{LO}}}^2/{\omega_{\mathrm{TO}}}^2$, using the experimental values of $\omega_{\mathrm{LO}}$ and $\omega_{\mathrm{TO}}$~\cite{fontana1990}.
 
%The large uncertainty reflects a large concentration of defects in the experimental samples.
% In addition to $C$, the optimized Curie temperature is $T_\theta=792$K. The temperature gap $\delta T=T_c-T_\theta$ is of $29$K, when using the value of $T_c=821$K extracted from the free energy studies discussed later in Section ~\ref{sec-cg}.  
% The simplest Landau theory for a first-order transition stipulates that the polarization-dependent free energy satisfies  $F(T,\mathcal{P})=A_0(T-T_\theta)\mathcal{P}^2-B_0\mathcal{P}^4+C_0\mathcal{P}^6$. From this expression one estimates a lower bound for phase coexistence given by $T_\theta$ and an upper bound given by $T^*=T_c+\delta T/3\approx 831$K. 

In a future publication~\cite{xie2024landau}, we will report the free energy difference between the ferroelectric and the paraelectric phase calculated with well-tempered metadynamics~\cite{barducci2008well} for a range of temperatures around $T_c$. Using metadynamics,
a well-established technique for enhanced statistical sampling, we can better estimate finite-size effects on $T_c$, confirming that the phase transition temperature of our model is $T_c=(821\pm 1)K$. Furthermore, the free energy profile as a function of the magnitude of the polarization indicates that the ferroelectric phase is metastable above $T_c$ and below $830K$, while the paraelectric phase is metastable roughly between $810$K and $T_c$. 
 
In summary, our DP molecular dynamics (DPMD) simulations describe the FE-PE transition of PbTi$\text{O}_3$ in good agreement with experiments. The predicted $T_c$ is $\sim 60$K higher than in the experiment. On the energy scale, this corresponds to $5$meV/atom, which is on the order of the statistical error of our energy model. 
We closely recover the experimental values of the tetragonality, enthalpy, and polarization changes across the first-order transition. The consistently good agreement with experiment found for the structural, thermodynamic, and dielectric properties probably results from the full inclusion of anharmonicity in our models, since the lattice distortion across the phase transition in PbTi$\text{O}_3$ can hardly be regarded as a small perturbation of a reference equilibrium structure. 

Having found that finite-size effects are almost negligible for $L\geq 12$, in the following, we will use $L=15$ for all results unless otherwise specified. 

\section{Emergence of strong orientational disorder}\label{sec-disorder}

%%%%%%%%%%%%%%%%%%%%%%%%%%%%%
\begin{figure}[th]
    \centering
    \includegraphics[width=\linewidth]{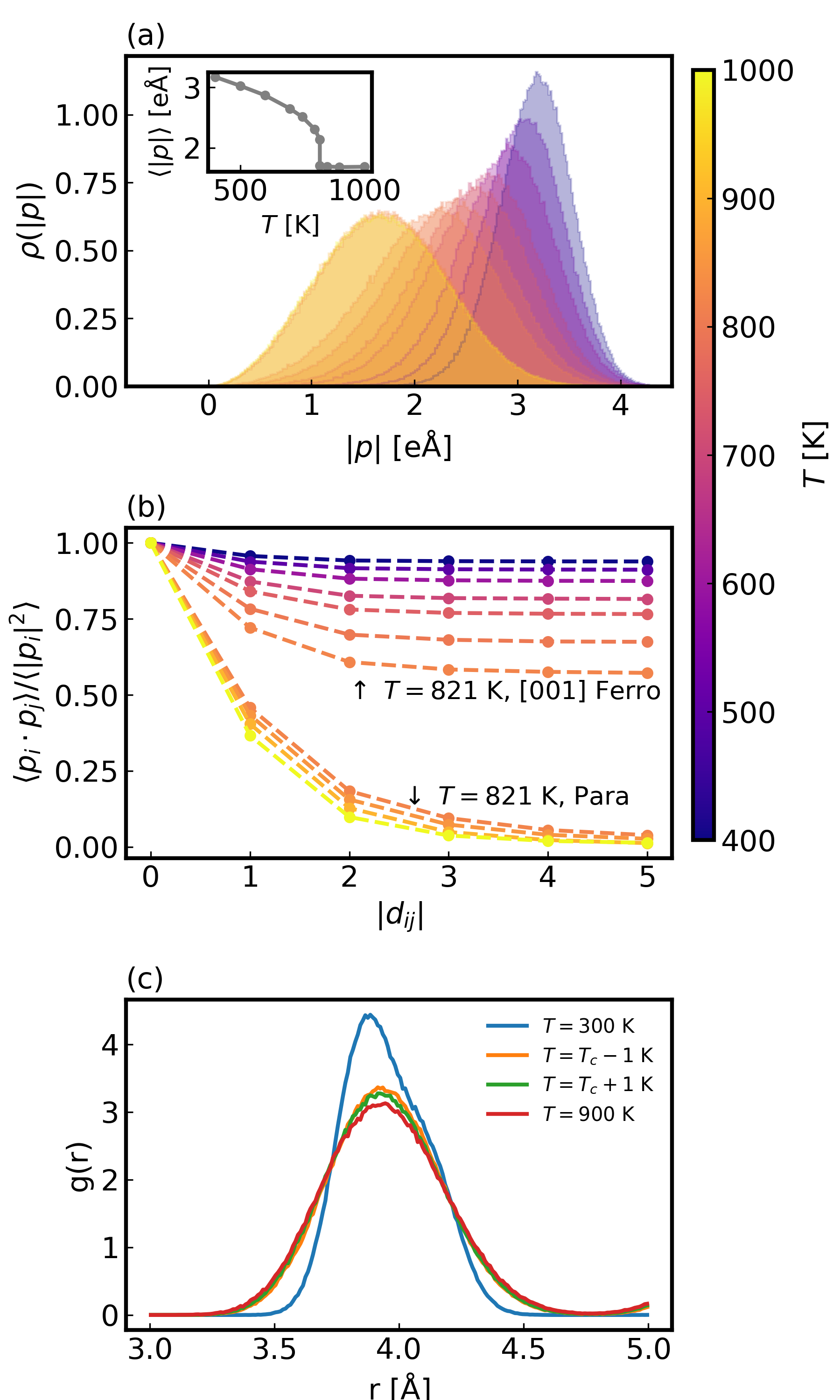}
    \caption{ (a) The PDF of $|p|$ at different temperatures. The PDF associated with lower temperature has a darker color, as indicated by the colorbar.
    The inset plots $\langle |p|\rangle$ versus $T$, showing a weak discontinuity around $T=T_c$. (b) Dipole-dipole pair correlation between unit cells spatially separated along [100] direction. $d_{ij}$ is the lattice vector from unitcell $i$ to $j$. The lines showing long-range order are associated with [001]-polarized ferroelectric phase. (c) The first peak in the Pb-Pb radial distribution function $g(r)$. %For $T=300$K, $g(r)$ peaks at $3.88\mathrm{\AA}$, almost the same as the lattice constant $a$. For other temperatures plotted, $g(r)$ peaks at around $3.94\mathrm{\AA}$.  
    }
    \label{dpdist}
\end{figure}
%%%%%%%%%%%%%%%%%%%%%%%%%%%%%

% %%%%%%%%%%%%%%%%%
%  \begin{figure}[t]
%     \centering
%     \includegraphics[width=0.8\linewidth]{Pb_RDF.png}
%     \caption{ The first peak in the Pb-Pb radial distribution function $g(r)$. For $T=300$K, $g(r)$ peaks at $3.88\mathrm{\AA}$, almost the same as the lattice constant $a$. For other temperatures plotted, $g(r)$ peaks at around $3.94\mathrm{\AA}$. }
%     \label{rdf}
% \end{figure}
% %%%%%%%%%%%%%%%%%%%%%%
% %%%%%%%%%%%%%%%%%
%  \begin{figure}[t]
%     \centering
%     \includegraphics[width=0.9\linewidth]{DipolePlane.png}
%     \caption{ The heatmap of the probability density of local dipole moments projected to the ($1\overline{1}0$) plane (left column) and the ($001$) plane (right column). The radius of the auxiliary circle is the average local dipole moment $\langle |p|\rangle$. }
%     \label{dpproj}
% \end{figure}
% %%%%%%%%%%%%%%%%%%%%%%
The first-order FE-PE transition is accompanied by strong disordering of the local dipoles. This is
illustrated in Fig.~\ref{dpdist}, which reports (a) the probability density distributions of $|p|$, the magnitude of the local dipole moment, and (b) the normalized dipole pair correlation functions, respectively, at temperatures ranging from $T=400$K to $T=1000$K. The probability density distributions $\rho(|p|)$ are close to Gaussian distributions. From $T=400$K to just below $T_c$, $\rho(|p|)$ decreases in the average dipole amplitude, $\langle |p| \rangle$, and increases in variance, indicating a gradual development of disorder. 
 % This is in qualitative agreement with the displacive to order-disorder crossover found below $T_c$ in experiments~\cite{fontana1990}. 
In the vicinity of $T_c=821$K,  $\langle |p|\rangle$ is $2.1\text{e\AA}$ in the ferroelectric phase and only slightly smaller ($1.7\text{e\AA}$) in the paraelectric phase, with a standard deviation of approximately $0.6\text{e\AA}$ in both cases. The reduction of $\langle |p|\rangle$ indicates a weak displacive effect across the phase transition, suggesting that the onset of the paraelectric phase should be mainly
due to increasing orientational disorder of the local dipoles with an ensuing loss of long-range order.  
This can be verified through the (normalized) dipole-dipole pair correlation function, shown in Fig.~\ref{dpdist} (b). In the paraelectric phase, the dipole-dipole correlations decay with the distance between the dipoles with a correlation length $\xi(T)<1$ nm that is largely insensitive to temperature. In the ferroelectric phase, long-range order is present and becomes stronger at lower temperature. The correlation function is discontinuous for $T=T_c$. At the same time, the first peak of the radial distribution function $g(r)$ (Fig.~\ref{dpdist}(c)) of the Pb atoms displays negligible discontinuity across $T_c$, indicating that the local geometric structure does not undergo major deformation across the transition. 

\begin{figure}[t]
\centering
\includegraphics[width=\linewidth]{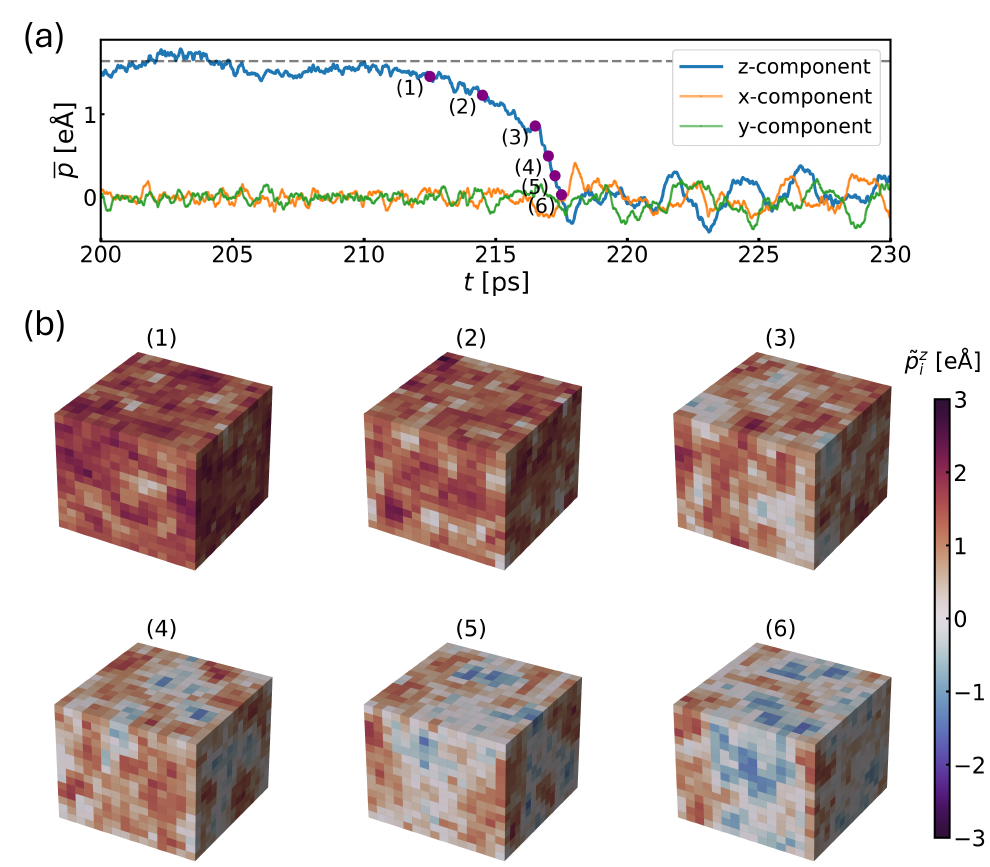}
\caption{ (a) The average local dipole moment $\overline{p}=p^G/L^3$ as a function of simulation time. The dashed grey line indicates the average $z$-component of the local dipole moment in the metastable ferroelectric phase. The six purple dots are labeled from (1) to (6), corresponding to the six local dipole configurations plotted below. (b) Typical local dipole configurations through the phase transition event. Local dipoles are represented by $15\times15\times15$ colored voxels. Each voxel is associated with an elementary cell. The color is mapped to $\tilde{p}_i^z$, indicated by the plotted colormap. }
\label{fig-trans}
\end{figure}
%%%%%%%%%%%%%%%%%%%%%%

%  %%%%%%%%%%%%%%%%%
% \begin{figure*}[t]
% \centering
% \includegraphics[width=\linewidth]{dipole_disorder.pdf}
% \caption{(a) The heatmap of the probability density of local dipole moments projected to the ($1\overline{1}0$) plane (left column) and the ($001$) plane (right column). The radius of the auxiliary circle is the average local dipole moment $\langle |p|\rangle$.  (b) The average local dipole moment $\overline{p}=p^G/L^3$ as a function of simulation time. The dashed grey line indicates the average z-component of the local dipole moment in the metastable ferroelectric phase. The six purple dots are labeled from (1) to (6), corresponding to the six local dipole configurations plotted below. (c) Typical local dipole configurations through the phase transition event. Local dipoles are represented by $15\times15\times15$ colored voxels. The color is mapped to $\tilde{p}_i^z$, indicated by the plotted colormap. }
% \label{fig-trans}
% \end{figure*}
% %%%%%%%%%%%%%%%%%%%%%%
The dynamic development of disorder can be followed in real time with non-equilibrium molecular dynamics. We simulate the FE-PE phase transition dynamics in a $L=15$ cell with a trajectory generated by NPT-MD at $T=T_c+2$K, 
starting from a ferroelectric configuration extracted from a well equilibrated trajectory at $T=815$K. The isothermal-isobaric condition is maintained using the MTK method ~\cite{martyna1994constant} (see Supplemental Material~\cite{SupplementalMaterial} for more details). 

We find that, after spending about 210 ps in the metastable ferroelectric phase, the system undergoes a rare-event transition that takes about 5 ps to complete. The evolution of the average local dipole moment
$\overline{p}=p^G/L^3$ in the vicinity of the phase transition is plotted in Fig.~\ref{fig-trans}(a). The corresponding configurations of the local dipoles are displayed in Fig.~\ref{fig-trans}(b), where each voxel is associated with a Ti-centered primitive cell with color coding representing the running average of $p_i^z(t)$, the local dipole moment in the $z$ direction, within a small time window spanning $0.6$ ps, i.e.,
$ \tilde{p_i}^z(t) = \frac{\int_0^{3\varsigma}  f(\tau) p_i(t-\tau)  d\tau}{\int_0^{3\varsigma}  f(\tau)}$.
Here, $f(\tau)=\exp (-\tau^2/2\varsigma^2)$ is a Gaussian filter with $\varsigma=0.2$ps. This choice of $\varsigma$ suppresses irrelevant fast vibrational modes with a frequency larger than the soft-mode frequency. Suppressing small and fast fluctuations makes it easier to identify ergodic polar/nonpolar nanoregions (NRs) with a lifetime on the order of the picosecond.

Examining the evolution of $\overline{p}$, we notice that the symmetry among its Cartesian components is restored upon the transition and that the fluctuations have a significantly larger magnitude in the paraelectric phase, consistent with the dielectric susceptibility depicted in Fig.~\ref{figmd}(f). 
% The large fluctuation is compatible with the dielectric anomaly displayed by static susceptibility (Fig.~\ref{figmd}(f)). 
These large fluctuations facilitate the reverse transition, paraelectric to ferroelectric, when the system is brought out of equilibrium at $T<T_c$.
%The vibrational pattern of the fluctuation is already described by Fig.~\ref{fig-corr}(c3), and we can identify the dominant contribution is the soft mode.
% Here one can infer the pattern of the paraelectric-ferroelectric transition event as a counterpart: starting from the metastable paraelectric phase, the large fluctuation induced by the soft mode will break ergodicity at some point, as a rare event, and the dominant direction of $\overline{p}$ will catastrophically evolve to the direction of spontaneous polarization. To verify, we simulate a paraelectric-ferroelectric transition event at $T=T_c-2$K and we find the evolution of $\overline{p}$ (not plotted here) is indeed similar to a reverse process of the one shown in Fig.~\ref{fig-trans} --- while one dominant component of $\overline{p}$ grows away from its typical value, the ergodic fluctuation along other directions is gradually suppressed.  
The local dipole configurations before the transition in Fig.~\ref{fig-trans}(b,1) show many nonpolar NRs, with volumes of the order of 1$\mathrm{nm}^3$, as nearly white voxels. Figs.~\ref{fig-trans}(b,2-4) depict configurations on approaching the transition, which show a growing number of $-z$-polarized NRs as blue voxels. Nonpolar and $-z$-polarized NRs appear as random fluctuations as opposed to stable nucleation sites. Figs.~\ref{fig-trans}(b,5-6) show the final stage of the transition, characterized by homogeneous stochastic excitation of NRs, each occupying a volume of the order of $1\mathrm{nm^3}$. 
%For neighboring opposite polarized NRs, the domain walls are not as sharp as those present at $T\ll T_c$ ~\cite{shin2007nucleation}, neither are they smooth as Neel or Bloch-type domain walls. 
We have also studied the evolution of the coarse-grained local dipole configurations in a reverse transition event, from the paraelectric phase to the ferroelectric phase. We found similar configurations in reverse temporal order and we do not plot them here. 

% %%%%%%%%%%%%%%%%%
%  \begin{figure*}[t]
%     \centering
%     \includegraphics[width=0.9\linewidth]{IR.png}
% \caption{ (a) The FIR absorption spectra of ferroelectric PbTi$\text{O}_3$. Grey dashed lines indicate the experimental phonon frequency (labeled by symmetry) determined by Raman spectroscopy at room temperature~\cite{foster1993anharmonicity}.  (b) The FIR absorption spectra of paraelectric PbTi$\text{O}_3$.}
%     \label{fig-ir}
% \end{figure*}
% %%%%%%%%%%%%%%%%%%%%%%

%%%%%%%%%%%%%%%%%
 \begin{figure*}[t]
    \centering
    \includegraphics[width=\linewidth]{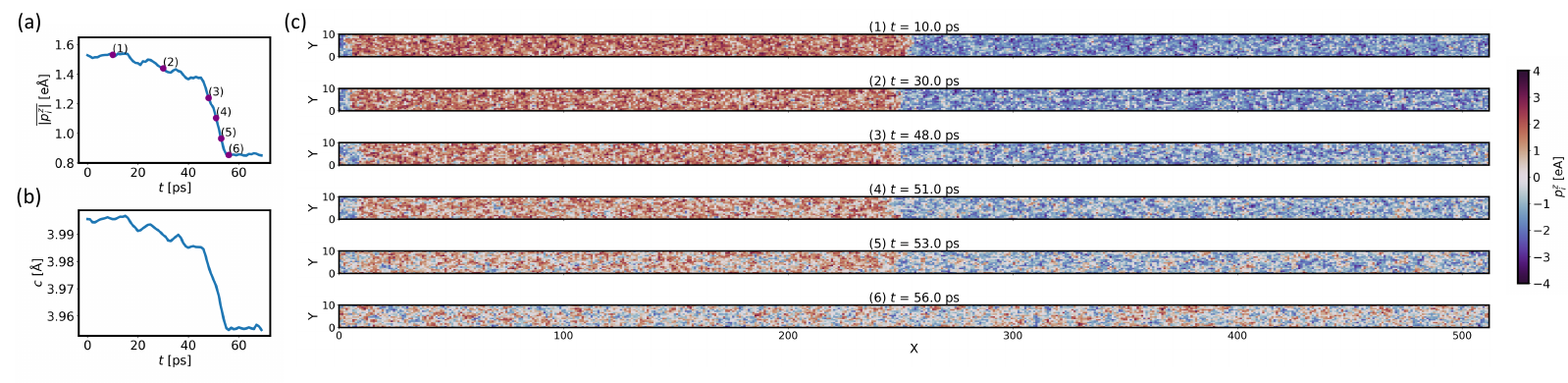}
\caption{ (a) Averaged absolute value of $p_z^i$ as a function of time, from MD simulation at $T=T_c+1$K. The six purple dots are labeled from (1) to (6), corresponding to the six local dipole configurations plotted in Panel (c). (b) The lattice constant $c$ as a function of time.  (c) Typical local dipole configurations of an arbitrary XY layer of the $512\times 16 \times 16$ simulation cell. Local dipoles are represented by $512\times16$ colored pixels. Each pixel is associated with an elementary cell. The color is mapped to ${p}_i^z$, indicated by the colormap. }
    \label{largercell}
\end{figure*}
%%%%%%%%%%%%%%%%%%%%%%

%%%%%%%%%%%%%%%%%
 \begin{figure*}[t]
    \centering
    \includegraphics[width=\linewidth]{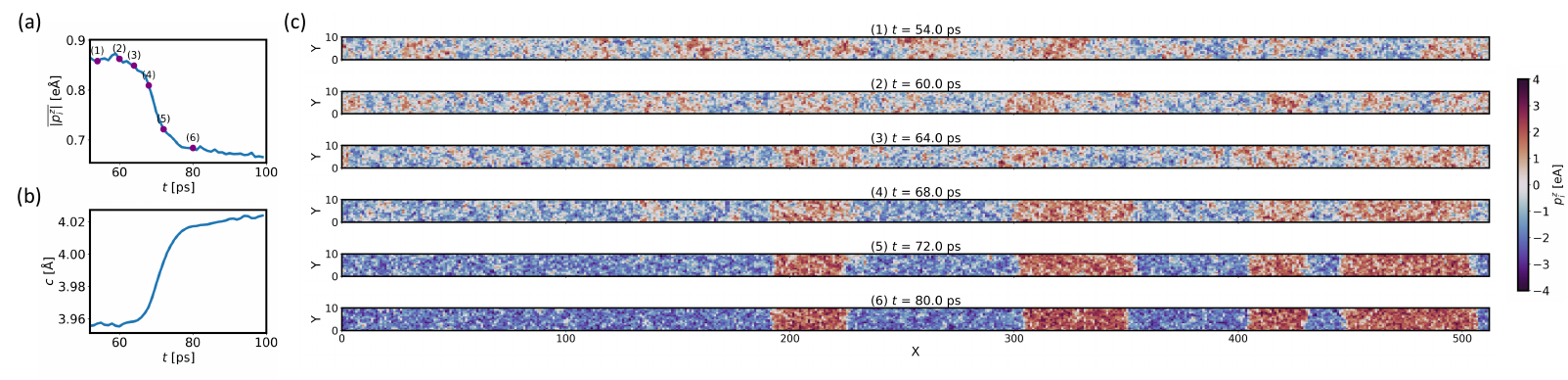}
\caption{ (a) Averaged absolute value of $p_z^i$ as a function of time, from MD simulation at $T=T_c-40$K.  (b) The lattice constant $c$ as a function of time.  (c) Typical local dipole configurations of an arbitrary XY layer of the $512\times 16 \times 16$ simulation cell. Pixel color is mapped to ${p}_i^z$, indicated by the colormap. }
    \label{overcool}
\end{figure*}
%%%%%%%%%%%%%%%%%%%%%%

The finding of a homogeneous phase transition without nucleation of the stable phase inside the metastable phase is puzzling, as the transition is first order and a free energy barrier separates the ferroelectric and paraelectric phases at $T=T_c$. Consistent with a homogeneous transition,
our study of the free energy surface with metadynamics
indicates that the barrier scales with volume~\cite{xie2024landau}.
By contrast, in typical first-order phase transitions, such as melting or ferroelectric switching by a driving field~\cite{shin2007nucleation}
the barrier scales with the surface leading to phase coexistence at the transition temperature. Surface scaling always occurs in models with short-range interactions, but the present model includes long-range interactions due to strain, which could originate a volume law for the barrier. However, we cannot exclude the possibility that the observed absence of nucleation and phase coexistence may be a mere effect of the finite size of the simulations. To test this hypothesis, we consider a long
periodic supercell $512\times 16 \times 16$
containing a 180-degree twin domain with opposite polarization along the z direction in the two halves of the box, as in panel (1) of Fig.~\ref{largercell}(c).

After equilibrating the system at $T=T_c+1$K for 50 ps with NPT-MD~\cite{martyna1994constant}(see Supplemental Material~\cite{SupplementalMaterial} for details), we set the temperature at $T=T_c+2$K and observe a transition from the twin-domain state to a paraelectric state, which took about 50 ps to complete, as indicated by the time evolution of the averaged absolute value of the local (cell) dipole component $p^i_z$ and of the average lattice constant $c$ shown in Figs.~\ref{largercell}(a)(b). If the transition was driven by nucleation, we expected it to initiate at the domain walls to minimize the energy cost. Instead, we observe the homogeneous transition depicted in Fig.~\ref{largercell}(c). Interestingly, the transition occurs in two stages. In the first stage (panels (1-3) of Fig.~\ref{largercell}(c)), the
twin domain remains
intact, apart from minor fluctuations in the walls, while the average magnitude of $p^i_z$ and the lattice constant $c$ decrease slightly. In the subsequent stage (panels (3-6) of Fig.~\ref{largercell}(c)) we observe a linear-in-time fast decay of $c$ and of the averaged absolute value of $p^i_z$ leading to a uniform paraelectric structure in less than 10 ps. Similar to Fig.~\ref{fig-trans} the transition occurs homogeneously over the sample. A homogeneous transition triggered by fluctuations at the nanoscale also occurs when an equilibrated paraelectric structure in a $512\times 16 \times 16$ supercell is abruptly undercooled to $T=T_c-40$K. The corresponding time evolution of the system is displayed in Figs.~\ref{overcool} (a)(b)(c). The final structure consists of four 180 degree twin domains with well-defined separating walls that are stable on the time scale of the simulation. The formation of twin domains reflects the stochastic nature of fluctuations at the nanoscale and suggests that the cost of an interface between domains of opposite polarization is small.

The lack of nucleation could be attributed to the strong elastic-dipolar coupling in lead titanate. Phase coexistence between sizable paraelectric and ferroelectric domains is hampered by the large mismatch between the cubic and the tetragonal lattices of the two phases. At coexistence, a  paraelectric domain applies compressive stress to the ferroelectric domain, leading to a long-range strain field and a volume law for the interfacial energy. In this situation, a homogeneous transition driven by nanoscale fluctuations may be favored over the coexistence of macroscopic domains, an outcome supported by our simulations and that can theoretically be justified with a simple phenomenological Landau model for the free energy per formula unit $g(\eta, \mathcal{P}, T)$ of a homogeneous domain with polarization $\mathcal{P}$ directed along z in the presence of a uniaxial strain field $\eta$ in the same direction.
The polarization is finite if the domain is ferroelectric and equal to zero if the domain is paraelectric. We ignore biaxial couplings between the z-polarization and the x / y-strain, as they are minor factors relative to the uniaxial coupling~\cite{Paul2017}.

Setting $T=T_c$ and omitting the temperature dependence in the following, the free energy density can be written
$g(\eta, \mathcal{P})=\frac{1}{2}B_{11} \eta^2 + K_{1zz} \eta \mathcal{P}^2 + f(\mathcal{P})$. Here, $B_{11}$ is the elastic constant corresponding to SCAN-DFT
($B_{11}\approx 132$eV~\cite{Paul2017}), $K_{1zz}<0$ is the elastic-dipolar coupling or electrostriction coefficient, $f(\mathcal{P})=\alpha_1\mathcal{P}^2+\alpha_{11}\mathcal{P}^4+\alpha_{111}\mathcal{P}^6$ is a sixth-order Landau-Devonshire expansion for the free energy density at the fixed lattice parameter $c_0$, which we take to be the equilibrium lattice parameter of the cubic paraelectric phase at $T_c$ ($c_0\approx 3.956\AA$). An arbitrary lattice parameter $c$ is related to strain $\eta$ through $c=(1+\eta)c_0$. The values of $c$ and $\mathcal{P}$ of the ferroelectric phase at $T_c$ are known from the MD simulation, which produces an equilibrium strain value $\eta_{\text{eq}}$ approximately equal to 0.01. 
Since $\eta_{\text{eq}}$ minimizes $g(\eta, \mathcal{P})$, we get the condition $B_{11}\eta_{\text{eq}} =-K_{1zz}\mathcal{P}^2$ from which the value of $K_{1zz}$ can be extracted. The free energy density of a paraelectric domain is recovered by setting $\mathcal{P}=0$ in the formula for $g(\eta, \mathcal{P})$.

The Landau model above can be used to estimate the energy cost of a macroscopic paraelectric domain that coexists with a macroscopic ferroelectric domain at $T=T_c$. We assume that the two domains retain their equilibrium bulk polarization, $\mathcal{P}$ and $0$, respectively, and that the domain walls are flat and perpendicular to the x-axis. The two domains share the same global strain $\eta$ under periodic boundary conditions, leading to an energy cost of the two-domain structure that scales with volume. At $T=T_c$ the two domains have the same free energy density, i.e. $g(\eta, \mathcal{P})=g(\eta, 0)$.  
Indicating by $x$ ($0<x<1$) the ratio between the volume of the paraelectric domain and the volume of the supercell, the free energy density of the two-domain structure is given by $g_{\text{fp}}(\eta,x)=\frac{1}{2}B_{11} \eta^2 + (1-x)K_{1zz} \eta\mathcal{P}^2 + (1-x)f(\mathcal{P})$, ignoring interfacial energies other than the elastic-dipolar contribution. Minimizing the free energy with respect to $\eta$ at fixed $x$ gives $\eta_{x}=-\frac{(1-x)K_{1zz}P^2}{B_{11}}$ for the equilibrium strain of the two-domain structure. After expressing
$K_{1zz}$ in terms of $B_{11}$, $\mathcal{P}$ and $\eta_{\text{eq}}$, this becomes $\eta_{x}=(1-x)\eta_{\text{eq}}$, where $\eta_{\text{eq}}$ is the strain associated with a single ferroelectric domain at equilibrium. Thus, the free-energy
density of the two-domain structure is $g_{\text{fp}}(\eta_{x},x)=\frac{1}{2}B_{11}\eta_{\text{eq}}^2(\mathcal{P})(x-x^2)$, which is minimal for $x=\frac{1}{2}$. The minimum free-energy density, $\Delta g = \frac{1}{8} B_{11} \eta_{\text{eq}}^2 \approx 1.65~\mathrm{meV}$ per formula unit, should be taken as an order of magnitude estimate of the barrier that separates the two coexisting phases at $T=T_c$, in view of the crude approximations made in the Landau model. In contrast, the barrier extracted from metadynamics simulations, $0.1~\mathrm{meV}$ per formula unit~\cite{xie2024landau}, 
is an order of magnitude smaller than $\Delta g$, suggesting that in the presence of long-range strain fields, the system may prefer to approach equilibrium homogeneously through fluctuations at the nanoscale, rather than through nucleation and growth of a macroscopic domain.
This conclusion is supported by our large-scale DPMD simulations exceeding the size of 100 nm in the long dimension. Sub-micrometer scales are relevant to most ferroelectric thin film devices, raising the possibility that a homogeneous transition driven by fluctuations at the nanoscale could be observed in epitaxial $\mathrm{PbTiO_3}$ in the absence of significant depolarization.  
However, we cannot exclude the possibility that the classical nucleation theory requiring a surface scaling barrier may be observed in larger samples. In the Landau model for the coupling of strain and polarization, we considered spatially uniform strain and polarization fields. A Landau-Ginzburg model with spatially varying fields would be more accurate and could lead to area law and phase coexistence in the thermodynamic limit.

\section{Displacive vs Debye relaxation dynamics}\label{sec-ir}

%%%%%%%%%%%%%%%%%
 \begin{figure*}[t]
    \centering
    \includegraphics[width=0.9\linewidth]{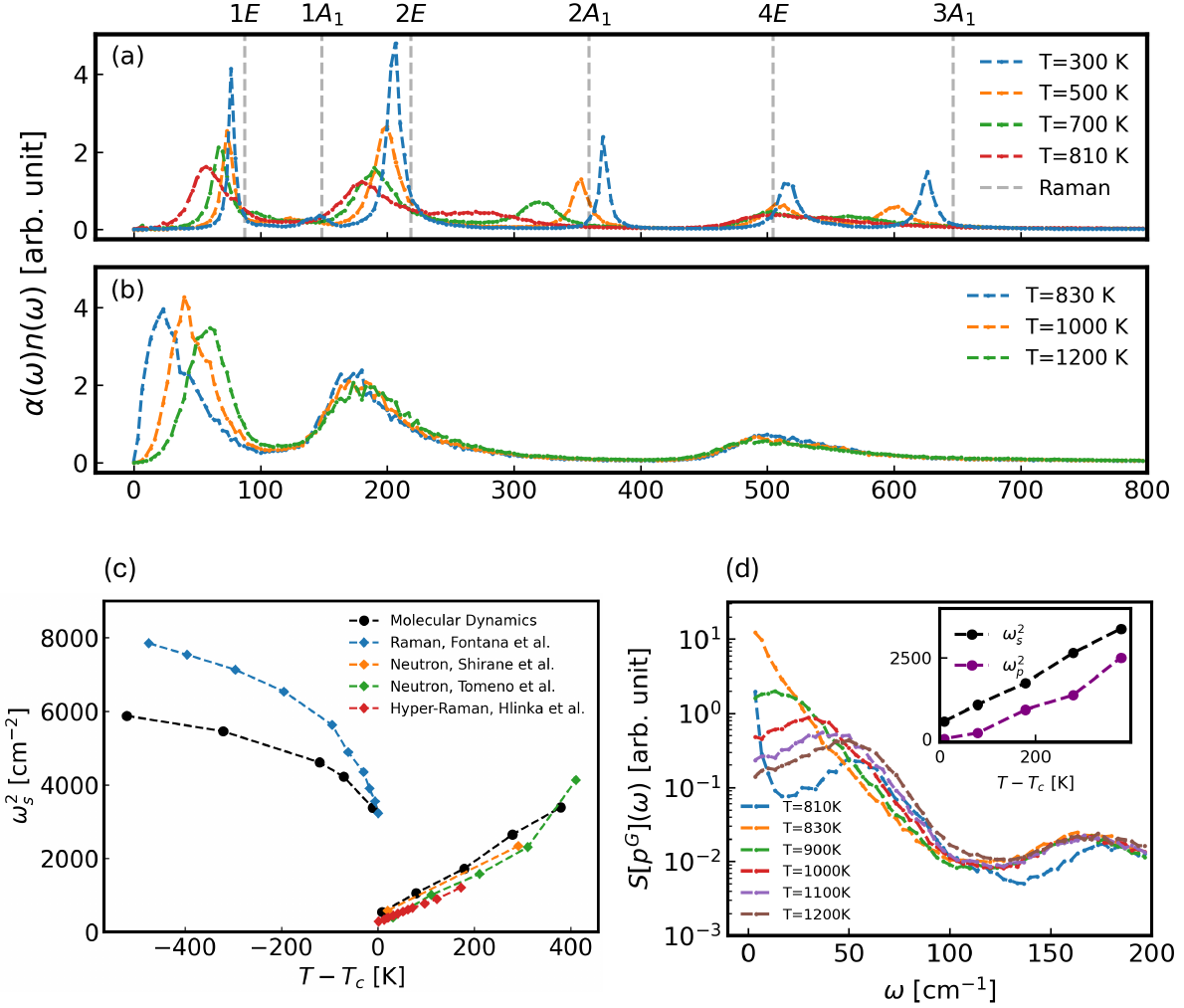}
\caption{ (a) The FIR absorption spectra of ferroelectric PbTi$\text{O}_3$. Grey dashed lines indicate the experimental phonon frequency (labeled by symmetry) determined by Raman spectroscopy at room temperature~\cite{foster1993anharmonicity}.  (b) The FIR absorption spectra of paraelectric PbTi$\text{O}_3$. (c) The ``soft mode'' frequency calculated by molecular dynamics compared to the experimental values~\cite{fontana1990, Shirane1970, Tomeno2012, Hlinka2013} as a function of the temperature deviation from $T_c$. (d) The power spectrum of the total dipole moment.  The inset plots the temperature dependence of $\omega^2_s$ and $\omega^2_p$ in the paraelectric state.
}
    \label{fig-ir}
\end{figure*}
%%%%%%%%%%%%%%%%%%%%%%

To detect putative features of displacive-type softening we compute the far-infrared (FIR) absorption spectrum of PbTi$\text{O}_3$. A previous calculation was limited to harmonic vibrational analysis~\cite{peperstraete2018}, neglecting anharmonicity, which is significant near the phase transition. 
Within linear response theory,  
% Here, we follow the MD approach proposed in Ref.~\cite{PhysRevB.102.041121} and simulate FIR absorption spectra based on the Fourier transform of the velocity autocorrelation function (ACF) of cell dipole $p^G$.
the product of the IR absorption coefficient per unit length, $\alpha(\omega)$, with the refractive index, $n(\omega)$, is given by the Fourier transform of the time autocorrelation function (ACF) of the time derivative of the total dipole moment $p^G$ via
\begin{equation}\label{eq:ir}
    \alpha(\omega)n(\omega) = \frac{2\pi \beta}{3cV}\int_{-\infty}^{\infty}  e^{-i\omega t}\langle \dot{p}^G(t)\cdot \dot{p}^G(0)\rangle dt.
\end{equation}
This approach, in combination with DP and DD models, was previously used in Ref.~\cite{PhysRevB.102.041121}.
Here, the results are given in Fig.~\ref{fig-ir}, for $T\in[300,1200]$K and $\omega \in [0,800]\mathrm{cm}^{-1}$. In this spectral range, the refractive index is either constant or smoothly varying with $\omega$. Hence the peaks in $\alpha(\omega)n(\omega)$ are due to absorption. 
To our knowledge, no experimental FIR absorption spectrum is available for comparison. Thus, in Fig.~\ref{fig-ir}(a), we report as vertical dashed lines the experimental frequencies measured by Raman spectroscopy at $T=300$K~\cite{foster1993anharmonicity}. They can be compared with the simulated FIR absorption spectrum at the same temperature reported in the figure as a dashed blue line.  
The phonon labels are as in~\cite{foster1993anharmonicity}. We identify all TO modes with the exception of $1B_1$ and $3E$, two modes split from the IR-inactive $B_1$ mode of the Pm$\overline{3}$m structure. The absence of $1B_1$ and $3E$ agrees with harmonic vibrational analysis~\cite{peperstraete2018}. Active modes $1E$, $2E$, and $4E$ are associated with dipolar vibration orthogonal to spontaneous polarization. Active modes $1A_1$, $2A_1$, and $3A_1$ are associated with dipolar vibration parallel to spontaneous polarization. Our simulation agrees well with the experiments for all these modes. The largest discrepancy occurs for the 3$A_1$ mode and amounts to only $20\mathrm{cm}^{-1}$. The calculated frequency of the soft 1$E$ mode, $77\mathrm{cm}^{-1}$ at $T=300$K, should be compared with an experimental value of $87.5\mathrm{cm}^{-1}$. Harmonic vibrational analysis~\cite{freire1988, peperstraete2018} shows that the $1E$ and $1A_1$ modes split from an imaginary frequency mode corresponding to a uniform displacement of the oxygen octahedron against the lead atoms in the Pm$\overline{3}$m structure.
% A naive displacive explanation of phase transition may then attribute the instability of the ferroelectric phase to the softening of $1A_1$ mode.  
% But here $1A_1$ mode appears with negligible magnitude at $300$K and becomes indistinguishable above, suggesting it plays no significant role in the transition.  

Fig.~\ref{fig-ir}(a) shows that all the phonons are increasingly softened and damped when $T_c$ is approached from below. The softening of all the phonons is expected from the significant temperature dependence of the lattice constants for $T<T_c$. Damping can be associated with an increase in disorder. In contrast, when $T_c$ is approached from above, only the lowest frequency mode is softened without significant overdamping, as shown in Fig.~\ref{fig-ir}(b). In Fig.~\ref{fig-ir}(c) we report the ``soft mode'' squared frequencies $\omega_s^2$ extracted from simulations at different temperatures, below and above $T_c$, and compare them to multiple experiments~\cite{fontana1990, Shirane1970, Tomeno2012, Hlinka2013}. In the ferroelectric phase, the simulated and the experimental $\omega_s^2$ vary approximately linearly with $T$ for $T<T_c-100$K, but strong deviations from linearity occur at higher temperatures, a behavior that was attributed to a crossover of the dominant phase transition mechanism from displacive to order-disorder~\cite{fontana1990}. Surprisingly,
the displacive behavior appears to be restored in the paraelectric phase, where calculated and experimental $\omega_s^2$ show a steady linear temperature dependence up to $T=T_c+400$K, raising the question of how an almost ideal ``soft mode'' can coexist with the strong disordering effects observed in the PE phase both in simulation and in diffraction experiments.
To answer this question, we observe
that $\alpha(\omega)n(\omega)$ in 
Eq.~(\ref{eq:ir}) is proportional to the product of $\omega^2$ times the spectrum of the ACF of the total dipole, $S[p^G](\omega)=\int_{-\infty}^{\infty}  e^{-i\omega t}\langle  p^G(t)\cdot p^G(0)\rangle dt$, as can be seen with integration by parts. $S[p^G](\omega)$
can detect low-frequency features that are suppressed by the $\omega^2$ factor in the IR absorption spectra. 

The spectra corresponding to $S[p^G](\omega)$ at different temperatures are plotted in Fig.~\ref{fig-ir}(d). When $T_c$ is approached from above, a prominent central feature emerges at zero frequency, merging with a broad ``soft mode'' feature at non-zero frequency. The central feature is associated with Debye relaxation due to dipolar disordering. The frequency of the broad peak, $\omega_P(T)$, is smaller than $\omega_s(T)$, the frequency of the ``soft mode'' in the IR spectrum (inset in Fig.~\ref{fig-ir}(d)),  
and vanishes near $T_c$, consistent with the dominant effect of disorder. The ``soft mode'' appears sharp in the IR spectra when $T$ approaches $T_c$ because the central component is strongly suppressed by the factor $\omega^2$. The lineshape of $S[p^G](\omega)$ and its evolution is not consistent with the damped harmonic oscillator models postulated in the quasi-harmonic theory of lattice vibrations.
The central peak in $S[p^G](\omega)$
is also present when $T_c$ is approached from below, as shown by the spectrum at $T=810$K in Fig.~\ref{fig-ir}(d). 
The presence of this central peak reflects the strong dipolar disorder that appears in the ferroelectric state as $T_c$ approaches, and is consistent with the rapid decay of the dipole pair correlations in the short range shown in Fig.~\ref{dpdist}(b). If the central peaks of $S[p^G](\omega)$ at $T=810$K and $830$K are attributed to Debye relaxation, the corresponding relaxation time $\tau_D$ can be extracted from the spectral function
$\omega S[p^G](\omega)$. This function is proportional to the imaginary part of the dielectric function and peaks at $\omega_D= \tau_D^{-1}$, the inverse of the Debye relaxation time. We find in this way that $\tau_D$ is between $0.8$ ps and $1.6$ ps at $T=830$K, while at $T=810$K, $\tau_D$ is at least $1.6$ps. We cannot estimate $\tau_D$ more precisely because our $\omega$ resolution is approximately $3.3\mathrm{cm^{-1}}$.
% indicating that disorder plays also an important role in the rapid decay of the dipole-dipole pair correlations shown in Fig.~\ref{dpdist}(b) for $T$ approaching $T_c$ from below. 
The central peak depicted in Fig.~\ref{fig-ir}(d), both above and below $T_c$, should be associated with the quasielastic feature detected with Raman scattering in Ref.~\cite{fontana1990}. 

The nearly ideal ``soft mode'' behavior of the experimental frequencies for $T>T_c$ reported in Fig.~\ref{fig-ir}(c) is likely due to the difficulty in resolving the quasi-elastic peak. Indeed, early inelastic neutron scattering experiments~\cite{Shirane1970} pointed out that the neutron group associated with the zone-center ``soft mode'' is rather broad and difficult to identify for $T<1000$K, so they derived the ideal displacive character with extrapolations based on measurements at higher temperature where the zone-center mode could be unambiguously identified. A similar ambiguity may
be present in Raman scattering experiments~\cite{Hlinka2013}, which displayed ``soft mode'' behavior using a damped harmonic oscillator model after empirically subtracting from the spectra the strong central peak due to elastic (Rayleigh) scattering. These considerations suggest that the postulated softening of a single zone-center optical phonon for $T>T_c$ is likely a secondary effect of a Debye relaxation driven by disorder. Our simulations show unambiguously
that the linear temperature dependence of $\omega^2_s(T)$ does not imply a uniform structural distortion in the transition to the paraeletric phase. 

\section{Conclusion}\label{sec:conclusion}
In this paper, we presented a comprehensive ab initio investigation of the phase transition of PbTi$\text{O}_3$. 
Our machine learning atomistic models fully include anharmonicity and describe macroscopic structural, thermodynamic, and dielectric properties in good agreement with experiments. 

With all-atom MD simulations, we find that strong disordering of the local dipoles plays a dominant role in driving a homogeneous FE-PE phase transition. The lack of a nucleation stage is found on sub-micrometer scales and is attributed to strong elastic-dipolar coupling. 
At the same time, by calculating the FIR absorption spectra,  we show that a non-overdamped ``soft mode'' is present in the paraelectric phase, whose square frequency ($\omega_s^2(T)$) exhibits linear temperature dependence in excellent agreement with experiments. We reconcile the strong disordering and the almost ideal displacive ``soft mode'' behavior for $T>T_c$ by associating the ``softening''  with a disordering-driven Debye relaxation that appears as a broad central component in the power spectrum of the polarization and strengthens as $T$ approaches $T_c$ from above. 
Remarkably, we also find a central peak when $T$ is closely below $T_c$, where it appears narrow, sharp, and distinct from the optical mode, in good agreement with the central peak identified experimentally ~\cite{fontana1990} to signify displacive to order-disorder crossover when $T$ approaches $T_c$ from below. 

It follows that the FE-PE transition of PbTi$\text{O}_3$ is dominated by strong disordering effects that give rise in the paraelectric phase to characteristic local dipole fluctuations that remind the ergodic relaxor state~\cite{bokov2020}, in which the polar nanoregions have ergodically fluctuating, instead of frozen, directions of local dipole moments. 

The same methods adopted here can be used to study the ferroelectric transition in other materials where disordering and soft-mode effects may coexist~\cite{yang2024deuteration}. %Through the development of this work, we have applied the same methods to the prototypical hydrogen-bonded ferroelectric material potassium dihydrogen phosphate~\cite{yang2024deuteration}, where new physical insights are also developed with accurate modeling of anharmonicity.

% Here we only provide a case study of a prototypical perovskite. The study of other prototypical members of ferroelectric materials, such as some hydrogen-bonded systems, is ongoing work. In addition, the DPMD approach may also be used to provide ab initio insight into the microscopic theories of ferroelectric relaxors. 

\section*{Data avaiability}
The datasets, models and scripts that support the findings of this study are publicly available on Github~\cite{githubdata}.

\section*{Acknowledgement}
We thank Karin M. Rabe, Linfeng Zhang, Bingjia Yang, Kehan Cai, Marcos Calegari Andrade, and Pablo Piaggi for fruitful discussions. 
% We also thank Han Wang for assistance in implementing the DeepMD Plumed Module. 
All authors were supported by the Computational Chemical Sciences Center: Chemistry in Solution and at Interfaces (CSI) funded by DOE Award DE-SC0019394. P.X., Y.C. and W.E were also supported by a gift from iFlytek to Princeton University. %ONR Grant N00014-13-1-0338.
Calculations were performed on the National Energy Research Scientific Computing Center (NERSC), a U.S. Department of Energy Office of Science User Facility operated under Contract No. DE-AC02-05CH11231. Calculations were also performed using Princeton Research Computing resources at Princeton University, which is a consortium of groups led by Princeton Institute for Computational Science and Engineering (PICSciE) and Office of Information Technology's Research Computing.

\bibliography{ferro}

\end{document}

\preprint{APS/123-QED}
\newcommand{\pto}{PbTi$\text{O}_3$ }
\newcommand{\bto}{BaTi$\text{O}_3$ }
\newcommand{\red}[1]{\textcolor{red}{#1}}
\newcommand{\blue}[1]{\textcolor{blue}{#1}}

% \title{Ab initio modeling of ferroelectrics phase transition: The case of \pto}
\title{Supplemental Material for ``Thermal disorder and phonon softening in the ferroelectric phase transition of lead titanate''}
 
\author{Pinchen Xie}
\author{Yixiao Chen}
\affiliation {Program in Applied and Computational Mathematics, Princeton University, Princeton, NJ 08544, USA}
\author{Weinan E}
\affiliation{AI for Science Institute, Beijing, China,\\
Center for Machine Learning Research and School of Mathematical Sciences, Peking University, Beijing, China}
\author{Roberto Car}
\affiliation {Department of Chemistry, Department of Physics, Program in Applied and Computational Mathematics, Princeton Materials Institute, Princeton University, Princeton, NJ 08544, USA}

\date{\today}

\maketitle

\section{SCAN-based static description of \pto}\label{app-dft}
    In our electronic structure calculations, we use norm-conserving pseudo-potentials (NCPP) ~\cite{hamann1979norm}. 
    Relative to approaches like PAW~\cite{blochl1994projector}, NCPPs require a much larger plane wave basis set for good convergence. This is not a major limitation, because we only need a finite set of several thousand static DFT calculations, instead of direct ab-initio MD simulations~\cite{PhysRevLett.55.2471}, to train the DP models. 
    Specifically, all self-consistent KS-DFT calculations are done with the open-source Quantum ESPRESSO v.6.7 code ~\cite{giannozzi2009quantum} with NCPPs from the SG15 database ~\cite{SCHLIPF201536}.  We include the semi-core 5d states of Pb and the semi-core 3s, 3p, 3d states of Ti into the valence. We adopt a kinetic energy cutoff of  $150$Ry for the plane-wave basis. In the self-consistent calculations for the primitive cell, $\Gamma$-centered $4\times4\times4$ Monkhorst-Pack grids are used for k-point sampling. For $3\times3\times3$ and larger supercells, we use $\Gamma$ point sampling only. With input from the self-consistent band structure calculations, the Wannier functions and the polarization are computed with the Wannier90 code ~\cite{pizzi2020wannier90} using $2\times2\times2$ Monkhorst-Pack grids. %The effective dipole and the polarization are further derived from the equivalence between the band polarization and the Wannier center. 

    Upon structural relaxation, the equilibrium cubic lattice constant of
    our \pto model with space group Pm$\overline{3}$m is $a=3.925$\r{A}. 
    For reference, the experimental value extrapolated to zero temperature is $\tilde{a}=3.93$\r{A}~\cite{Mabud1979}.  The equilibrium tetragonal lattice constants with space group P4mm are $a=3.846$\r{A} and $c=4.393$\r{A}, respectively, corresponding to a tetragonality $c/a=1.142$. The off-centering displacement (in units of the lattice constant $c$) of titanium is $\Delta_{Ti}=0.049$. The displacement of oxygen is $\Delta_{O_1}=0.151$ and $\Delta_{O_2}=0.147$. The energy difference between the equilibrium Pm$\overline{3}$m phase and the P4mm phase is $\Delta E=26.9$meV/atom. 

    SCAN-based \pto suffers from the super-tetragonality problem, i.e. $c/a$ was overestimated compared to the extrapolated experimental tetragonality  $1.071$~\cite{Mabud1979}. At the same time, $\Delta_{Ti}$, $\Delta_{O_1}$ and $\Delta_{O_2}$ are all overestimated by $20\%\sim 30\%$ compared to the experimental measurements~\cite{Shirane:a01619}. 
    %Though far less severe than PBE-based models, the super-tetragonality indicated the SCAN-based model still lacked the subtle accuracy in balancing two inequivalent Ti-O bonds along the polarization direction, which are crucial for nearly all measurable properties of \pto. 
    To quantify the subtlety of tetragonality, we compute the potential energy of the relaxed tetragonal structure with a cell fixed to the experimental value and find it to be only $1.8$meV/atom higher than the one with a variable cell. This energy difference is much smaller than chemical accuracy, not to mention the inherent error of meta-GGA. %Instead of we adopt ad hoc adjustments to compensate for this issue in the main text.  
    
    % too small to be corrected within meta-generalized gradient approximations GGA   semi-local density functional. The inherent error of practical KS-DFT methods is almost always much larger than chemical accuracy, which is already tens of meV per atom. The partial success of SCAN largely depends on error cancellation, something that can not be systematically controlled or predicted. Beyond DFT, quantum chemical methods such as coupled-cluster-based ones are simply too expensive, if not impossible, for our disposal. Within the near future, one would expect the sensitive super-tetragonality can not be corrected in a systematic ab initio way for large scale numerical applications. So ad hoc adjustments should be considered if one cares about anything beyond qualitative agreement. However, to keep the generality of first-principles modeling, the ad hoc adjustments must not intrude the essence of DFT, otherwise the altered approach will degrade to special-purpose empirical methods. We discuss the proper non-invasive ad hoc adjustment in the main text.

    With KS-DFT results, we further compute the maximally localized Wannier functions and the associated Wannier centers for all valence bands. %The equivalence between the Wannier centers and the Berry phase polarization yields the cell polarization up to a quantum~\cite{Resta2007}. %But note that the polarization obtained this way is an approximation to the many-body polarization when periodic boundary condition is imposed~\cite{PhysRevLett.78.294}. 
    The polarization we obtained for the equilibrium tetragonal P4mm structure as opposed to the Pm$\overline{3}$m structure is $111\mu\text{C}/\text{cm}^2$. For the primitive cell, we obtain 22 MLWCs as shown in Fig.1 of main text.  Within the scope of this work, Pb atom always has six MLWCs. Ti and O always have four. So the Wannier centroid of an atom is defined without ambiguity. The effective charge of the Wannier centroid is the sum of the charges of the MLWCs. For the equilibrium P4mm structure, the Wannier centroid of $\text{O}_1$ is displaced from its home atom by $0.102$\r{A}. The Wannier centroid of $\text{O}_2$ is displaced from its home atom by only $0.008$\r{A}. This is in agreement with the previous observation that the displaced Ti redistributes the electron density along the O-Ti-O chain~\cite{marzari1998maximally} for BaTi$\text{O}_3$. But here Pb also play roles in the hybridization mechanism.

\section{Model Training}\label{app-model}
    Learning atomistic models from KS-DFT consists of several steps: data design, data generation and model training. By now these procedures have more or less become standard. For the rest of this section, we will try to describe these procedures for ferroelectrics without the technical details that have already been mentioned elsewhere~\cite{PhysRevLett.120.143001,ZHANG2020107206, PhysRevB.102.041121}. 

    \subsection{Data Design}
        
        First, we describe the format of ab initio data for the two DP models. Each data point consists of the atomic configuration and associated physical quantities.
        For a given \pto configuration in a supercell with periodic boundary conditions, the labels for training the energy model are the adiabatic potential energy, the Hellmann-Feynman forces and the virial tensor computed from KS-DFT. The labels for training the dipole model are the corresponding local dipole moments and the total dipole moment.    

        We use the $3\times 3\times 3$ supercell (135 atoms) in KS-DFT calculations to generate the training data for the two DP models --- they are both short range with the cutoff radius of $6$\r{A}.  %considering the distance between titanium and the third oxygen shell is around $6$\r{A} in all our simulations.  
        The short-range approximation adopted by our energy model is adequate for \pto because the long-range electrostatic interactions are treated correctly in the KS-DFT data. It will be effectively included in the trained energy model applied to the periodic structure, especially for the contribution from soft modes and long wave-length acoustic modes. For the same purpose, the effective Hamiltonian methods include long-range dipole-dipole interactions in addition to short-range coupling. What may not be captured by our short-range model is the non-analytic behavior of the dynamical matrix near the zone center which affects the LO modes, which are likely only a minor factor in describing the phase transition.

    \subsection{Data Generation and Training}
        We are interested in the property of \pto within $T\in [300, 1200]$K and $P\in[0, 10^5]$Pa. The data within this thermodynamic range are collected with the active learning procedure introduced in ~\cite{zhang2019active}. We use the DP-GEN~\cite{ZHANG2020107206} code to automate this procedure, the LAMMPS~\cite{plimpton1995fast} code as the MD engine and the DeePMD-kit code~\cite{wang2018deepmd, lu2021dp} to train DP models.
        %%%%%%%%%%%%%%%%%%%%%%%%%%%%%%%%%%%%%%%%
        \begin{figure}[tb]
            \centering
            \includegraphics[width=0.7\linewidth]{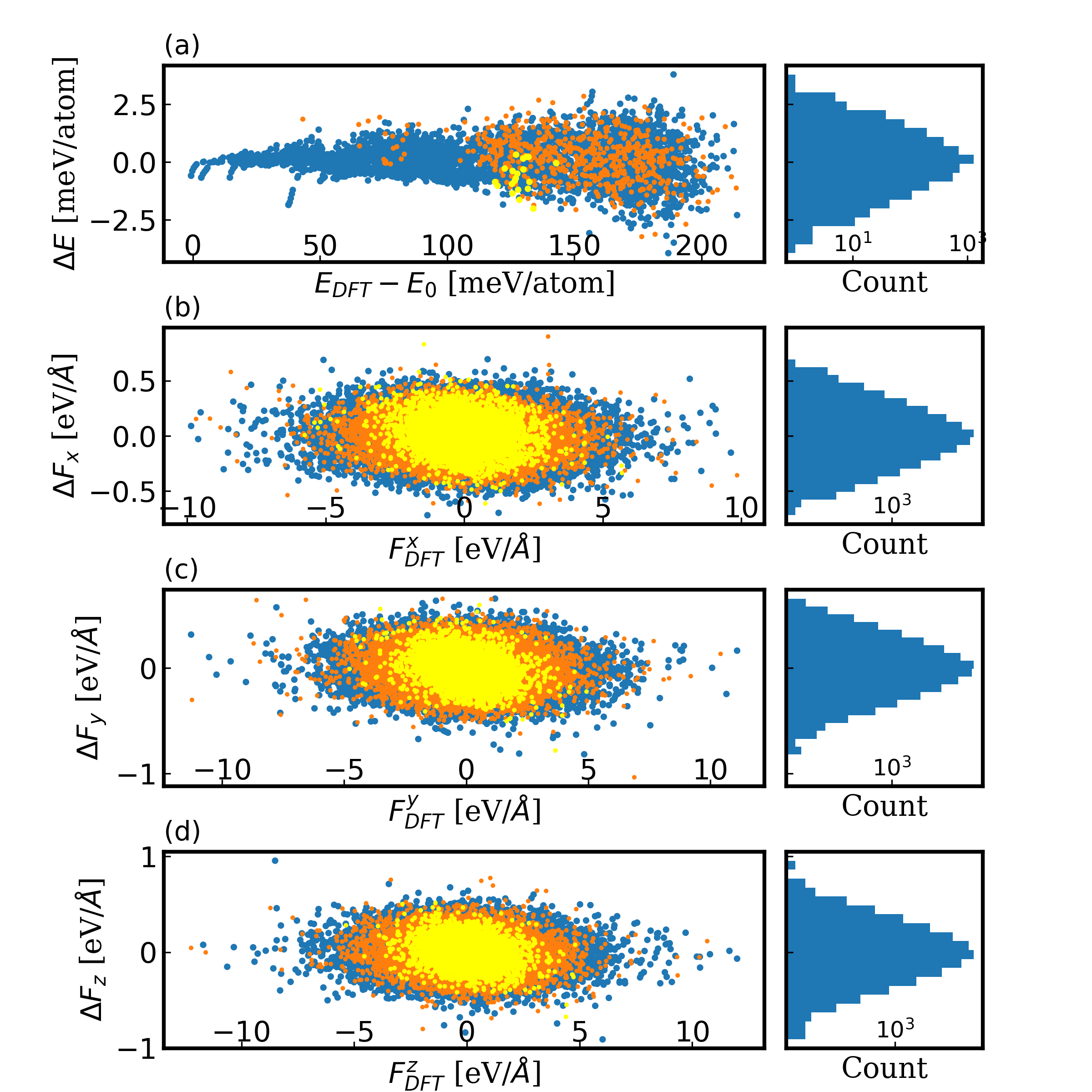}
            \caption{Error distribution of the energy model on the training set (Blue plots), test set (orange plots) and extra test set (yellow plots). $E_{\text{DFT}},F^{x,y,z}_{\text{DFT}}$ are the energy and force labels from the DFT data. $\Delta E, \Delta F_{x,y,z}$ are the difference between the model prediction and the label. $E_0$ is a constant used to shift the plot. (a) Left: The distribution of $\Delta E$ with respect to $E_{\text{DFT}}$. Right: The histogram of $\Delta E$ for the training set. (b-d) Left: The distribution of $\Delta F_{x,y,z}$ with respect to $F^{x,y,z}_{\text{DFT}}$. Right: The histogram of $\Delta F_{x,y,z}$ for the training set.}
            \label{dp-train-error}
        \end{figure}
        %%%%%%%%%%%%%%%%%%%%%%%%%%%%%%%%%%%%%%%%
        %%%%%%%%%%%%%%%%%%%%%%%%%%%%%%%%%%%%%%%%
        \begin{figure}[tb]
            \centering
            \includegraphics[width=0.7\linewidth]{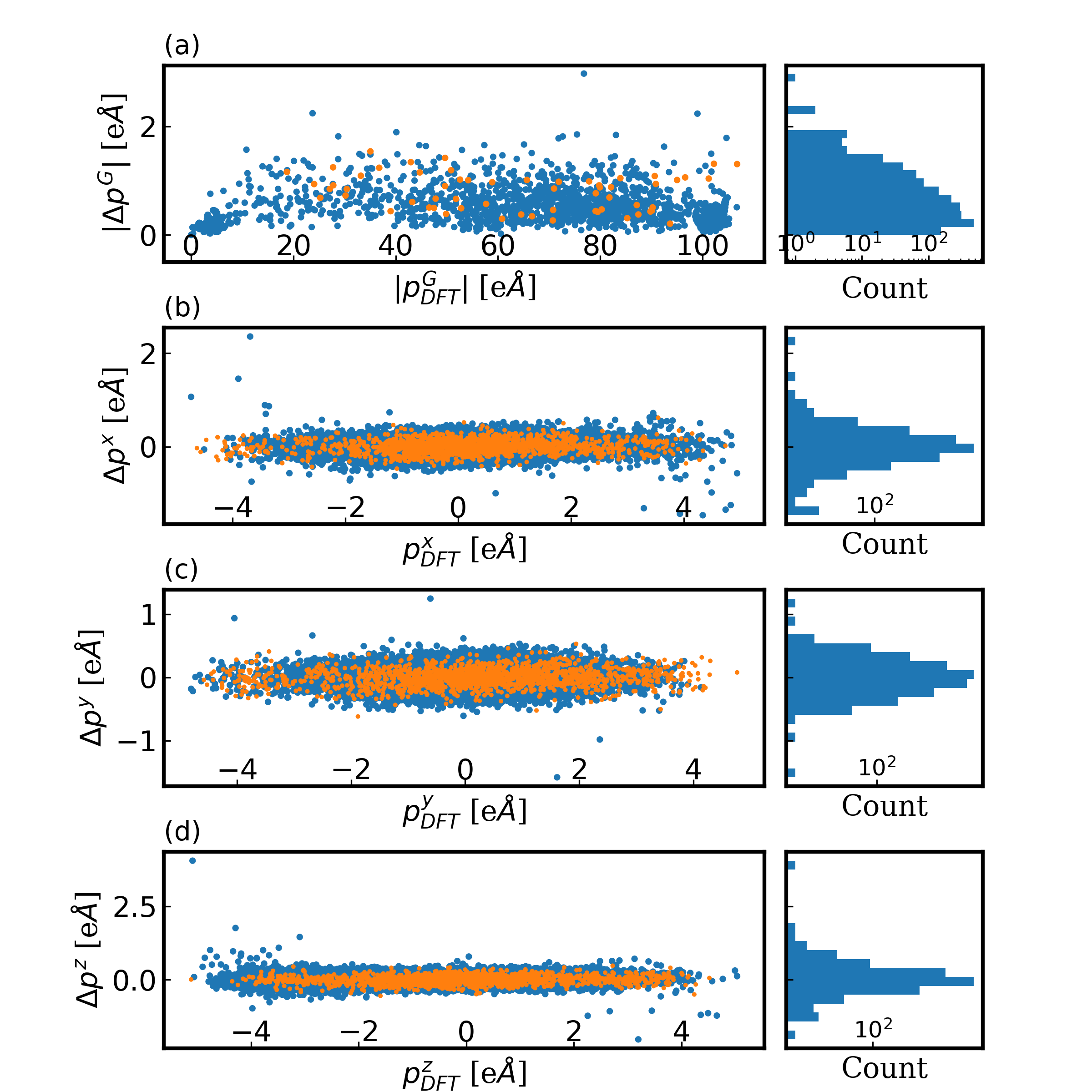}
            \caption{Error distribution of the dipole model on the training set (Blue plots) and test set (orange plots). $p^G_{\text{DFT}},p^{x,y,z}_{\text{DFT}}$ are the global and local dipole labels from the DFT data. $\Delta p^G, \Delta p^{x,y,z}$ are the difference between the model prediction and the label. (a) Left: The distribution of the 2-norm of $\Delta p^G$ with respect to the 2-norm of $p^G_{\text{DFT}}$. Right: The histogram of $|\Delta p^G|$ for the training set. (b-d) Left: The distribution of $\Delta p^{x,y,z}$ with respect to $p^{x,y,z}_{\text{DFT}}$. Right: The histogram of $\Delta p^{x,y,z}$ for the training set.}
            \label{dipole-train-error}
        \end{figure}
        At the end of the active learning procedure, we collect 5032 data points together with the energy, force, and virial labels.
        4432 data points are used to train the energy model. The other 600 data points are used for validation.

        Supplemental Figs.~\ref{dp-train-error} (blue and orange plots) show the prediction accuracy of the energy model compared to the DFT data. 
        The distribution of the error is roughly Gaussian. For energy prediction, the root mean square error (RMSE) is around $0.7$meV per atom for the training set and $1.0$meV per atom for the test set. For the force prediction, the RMSE is around $0.29\mathrm{eV/\AA}$ for the training set and $0.35\mathrm{eV/\AA}$ for the test set. Hence, the energy model is faithful to the current dataset. All DFT data for training used the $3\times 3 \times 3$ supercell. To consider the generality of the model, we should test the model with DFT data with different supercells. 
        %However, the generality of the trained short-range model is limited by its finite cutoff and the finite periodic boundary condition of the data. Specifically, the $3\times 3 \times 3$ supercell used to produce the data provides the short-range model with the configurations that the neighboring atoms outside the cutoff radius is the replicate of those inside the cutoff radius. So the long-range dipole-dipole interaction are modeled under such a periodic environment. To quantify the error, we had to compare the prediction of the energy model with SCAN DFT results on atomic configurations with a different periodicity.  
        To this end, we generated an additional test set consisting of 20 supercell atomic configurations $4\times 4 \times 4$, collected at NPT-MD simulations in the tetragonal phase. The error made by the energy model on this data set is shown in the yellow plots in the Supplemental Fig.~\ref{dp-train-error}. The error in the energy does not show a tendency to increase. The error in the force assumes a similar Gaussian distribution as in the previous data set.  Thus, we conclude that our short-range energy model is faithful to the SCAN-based KS-DFT inside the temperature and pressure range specified by our dataset, with a deviation much smaller than the threshold of chemical accuracy. %But note that this accuracy only applies to bulk configurations without defects, boundary effects or strong electric fields unless additional data are added to the training set.

        In addition, we calculate the optimal lattice constants for structures with space group P4mm and Pm$\overline{3}$m respectively. The cubic lattice constant is $a_{\text{DP}}=3.93$\r{A}, the same as the SCAN-DFT result. The tetragonal lattice constants are $a_{\text{DP}}=b_{\text{DP}}=3.86$\r{A} and $c_{\text{DP}}=4.30$\r{A}, slightly different from the SCAN-DFT results $a_{\text{DFT}}=b_{\text{DFT}}=3.846$\r{A} and $c_{\text{DFT}}=4.393$\r{A}. Further analysis shows that the energy model yields $0.6$meV$/$atom difference between these two tetragonal structures while SCAN-DFT yields $1$meV$/$atom difference. This deviation is compatible with the error distribution of the energy model.

    % \subsection{Dipole model}

        Dipole labels are added to the data set after training the energy model. We computed the dipole labels for only part of the dataset because the entire dataset contained redundancy. In addition, the generation of dipole labels is much more expensive than the others. To determine which data point should be labeled, we train an ensemble of energy models with different reduced training sets. Then we compare the models trained with the reduced datasets to the productive energy model trained with the entire training set in terms of error distribution and structural relaxation. It turns out that a reduced dataset containing 1835 data points is already enough to produce an energy model with basically the same level of accuracy as the production model. This is expected since the initial dataset contains a lot of similar atomic configurations from very short ab initio MD trajectories. In addition, a lot of data points generated in the early stage of the learning process became redundant in the final data set. %We also benefitted from the fact that we were modeling only two solid phases of \pto without defects where local chemical environment is simpler than liquids.

        We generate dipole labels for the reduced training set consisting of 1835 data points. In addition, 
        we generate a test set consisting of 61 data points collected using NPT-MD simulations in both cubic and tetragonal phases.
        The error distribution of the final dipole model is shown in Supplemental Fig.~\ref{dipole-train-error}. For the global dipole prediction associated with supercells $3\times 3 \times 3$, the RMSE is around $1.1\mathrm{e\AA}$ on the training set and $1.5\mathrm{e\AA}$ on the test set. In terms of polarization, the RMSE is $1.1\mathrm{\mu C/cm^2}$ in the training set and $1.4\mathrm{\mu C/cm^2}$.
        For local dipole prediction, the RMSE is around $0.3\mathrm{e\AA}$ in the training set and $0.4\mathrm{e\AA}$ in the test set. 
         %Meanwhile,  the local dipole prediction is slightly worse. 
        % Within the scope of this work, we need only high accuracy on the global dipole which is rigorously defined by the modern theory of polarization. The local dipoles here are merely auxiliary variables. 
        The results suggest that the dipole model can accurately predict the polarization change caused by structural distortion. %So we will disregard the slight inaccuracy in its prediction and proceed with this dipole model.  
         
        For comparison, we also fit a linear model with static Born charges as trainable parameters to all our dipole data. The RMSE of the linear model on global dipole data is twice as large as our dipole model. For configurations near the two limit $|p_{\text{DFT}}^G|\approx 0$ and $|p_{\text{DFT}}^G|\approx 100e$\r{A}, the linear model gives outliers with errors approximately three times the RMSE. This implies that the trained parameters effectively take the average of the cubic-phase Born charges and the tetragonal-phase Born charges. %That being said, the overall performance of the linear model is acceptable for finite temperature modelling if one do not pursue uniform convergence towards the first-principles results. When the fitted linear model is used for enhanced sampling, it predicts the same phase transition temperature for the energy model as the neural network dipole model. This is expected since any order parameter signifying the displacive feature of the geometric phase transition considered here should be proper. %For future studies along the same path, we suggest that the linear model fitted to fisrt-principles data can be a good substitution for the neural network dipole model when the dipole data are too expensive to acquire or the computational resources can not support enhanced molecular dynamics with two deep neural networks at the same time. 

\section{Technical Details of MD simulation}\label{app-md}

     MD simulations were carried out with the joint use of the DeePMD kit, LAMMPS, and PLUMED~\cite{bonomi2019promoting, tribello2014plumed} with an additional package~\cite{yixiao} that implements the dipole model as collective variables.
    
    The results on lattice constants, enthalpy, spontaneous polarization, specific heat, dielectric susceptibility, and distribution/correlation of local dipole moments were obtained by MD simulations with a time step $\Delta t=0.5$ fs and periodic boundary conditions. 
    The isothermal-isobaric condition was maintained with the MTK method ~\cite{martyna1994constant} using the default parameters of LAMMPS. The total time length of each NPT-MD simulation was approximately 1 ns. This was sufficient for computing equilibrium averages, starting from an optimized tetragonal atomic configuration for $T \leq 821$K, and from an optimized cubic atomic configuration for $T > 821$ K.

    The results on phase transition dynamics of the $15\times 15\times 15$ supercell and the FIR absorption spectra were obtained with unbiased MD simulations using the time step $\Delta t=0.5$fs and the same type of thermostat and barostat. The simulation on phase transition dynamics of the $512\times 16\times 16$ supercell uses a large time step $\Delta t =2$fs while keeping other setups the same. 
    The damping times of the thermostat and barostat were set at $5$ ps. Each FIR absorption spectrum for $T<T_c$ was extracted from a 4ns trajectory of an L=12 supercell.
    Each FIR absorption spectrum for $T>T_c$ was extracted from a 2ns trajectory of an L=15 supercell. The larger supercell was used in the paraelectric phase to better reduce finite-size effects.

\bibliography{ferro}